\documentclass[12pt]{iopart}

\usepackage{bbm}
\usepackage{mathrsfs}
\expandafter\let\csname equation*\endcsname\relax
\expandafter\let\csname endequation*\endcsname\relax
\usepackage{amsmath}
\usepackage{hyperref}
\usepackage{color} 
\usepackage{ulem}

\usepackage{graphicx}
\usepackage{subfigure}

\usepackage{makecell}
\graphicspath{{Figures/}}

\begin{document}

\title{Relaxation of the entanglement spectrum in quench dynamics of topological systems}
\author{Yi-Hao Jhu$^1$, Pochung Chen$^1$ and Ming-Chiang Chung$^{2,3}$ }
\address{$^1$ Physics Department, National Tsing Hua University, Hsinchu, 30013, Taiwan}
\address{$^2$ Physics Division, National Center for Theoretical Science, Hsinchu, 30013, Taiwan}
\address{$^3$ Department of Physics, National Chung Hsing University, Taichung, 40227, Taiwan}
\ead{mingchiangha@phys.nchu.edu.tw}

\begin{abstract}
We study how the entanglement spectrum relaxes to its steady state in one-dimensional quadratic systems after a quantum quench. In particular we apply the saddle point expansion to the dimerized chains and 1-D $p$-wave superconductors. We find that the entanglement spectrum always exhibits a power-law relaxation superimposed with oscillations at certain characteristic angular frequencies. For the dimerized chains, we find that the exponent $\nu$ of the power-law decay is always $3/2$. For 1-D $p$-wave superconductors, however, we find that depending on the  initial and final Hamiltonian, the exponent $\nu$ can take value from a limited list of values. The smallest possible value is $\nu=1/2$, which leads to a very slow convergence to its steady state value.
\end{abstract}

\vspace{2pc}

\maketitle

\section{Introduction} \label{sec.I}

The experimental discovery of new topological phases such as topological insulators\cite{Konig2007,Roth2009,Bernevig2006}, topological superconductors\cite{Levy2013,Sasaki2011,Mourik2012} and Weyl semimetals\cite{Xu2015,Xu2015a,Lv2015,Lv2015a,Yang2015,Huang2015} symbolizes a new era of physics. Different from the normal ordered phases characterized by the broken symmetry, the topological phases  can be identified by either the Chern number or zeros of the Pfaffian related to Berry phase according to the symmetries\cite{Qi2011,Schnyder2008}. An important feature of topological phases is  the possibility to create the nonlocal correlations among subsystems of the quantum matter, which can be measured by the entanglement between the subsystem and the environment\cite{Kitaev2006,Levin2006,Li2008,Pollmann2010,Turner2011}. 
An useful entanglement measurement is the von Neumann entropy of a subsystem $A$ defined as $S_A = -\mathrm{Tr} \rho_A \log \rho_A$, where $\rho_A = \mathrm{Tr} \mid \Psi_{A \cup B} \rangle \langle\Psi_{A\cup B} \mid$ is the reduced density matrix after tracing out the environment $B$ from the whole system $A \cup B$\cite{Eisert2010,Amico2008}. Taking the form of entropy, entanglement entropy $S_A$ is a real entanglement measurement: under local operations and classical communications (LOCC)  a copy of the  pure state $\mid \Psi_{A\cup B} \rangle$  can be transformed into a state with the same or lower entanglement qualified by the entanglement entropy $S_A$. Furthermore, one can define the entanglement Hamiltonian ${\cal H}_E $ via the relation $\rho_A = e^{-{\cal H}_E } / {\mathrm{Tr} e^{-{\cal H}_E }}$\cite{Peschel2003}. Actually, entanglement Hamiltonian contains more informations than the entanglement entropy. For instance, it can be used to detect the topological order\cite{Kitaev2006,Levin2006}. For topological systems, edge states will appear when the topological system is measured by the entanglement spectrum, which is defined as the eigenvalues of the entanglement Hamiltonian ${\cal H}_E$\cite{Ryu2006,Qi2012}. This phenomenon can be viewed as the bulk-edge correspondence of entanglement measurement\cite{Chandran2011}. 

On the other hand, the thermodynamics of time evolved manybody systems has attracted a lot of attentions due to a number of ground-breaking experiments with ultracold atoms\cite{Bloch2008,Yukalov2011,Polkovnikov2011}. 
Historically speaking, either from the classical or quantum mechanics point of view, the statistical average of an observable  should be taken as the infinite-time average. However, the concept of an ensemble simplifies the calculation. In the statistical mechanics, it is generally believed that there exists an time-independent ensemble, where the ensemble average is equal to the infinite-time average. For example, a microcanonical ensemble takes all points of equal energy in the phase space with the same probability. The ergodicity of the system ensures that the microcanonical ensemble average of observables is equivalent to the time average. Furthermore, with a heat reservoir keeping a constant temperature, the microcanonical ensemble for the system plus the reservoir ensures the canonical ensemble for the system. Therefore,  the ergodicity is the key ingredient of the microcanonical and canonical ensembles.

It is known that non-integrable models, which are in general ergodic, thermalize in the long run\cite{DAlessio2016}. Integrable models, however, lack the ergodicity due to certain constant equations of motion. Therefore for integrable models, the (micro)canonical ensemble can not describe the time average of observables. It was first conjectured by Rigol {\it et. al.} \cite{Rigol2007,Rigol2006}  that the asymptotic steady state of a quantum quench is described by the generalized Gibbs ensemble (GGE)\cite{Essler2016}. The density matrix of GGE is obtained as the less biased guess of the steady state given the constraints on the dynamics stemming from  the existing set of nontrivial integrals (that is also why the ergodicity fails.)  Till now GGE is still under debate. However, for the general Gaussian initial states, it has been proven by Cazalilla, Iucci and Chung \cite{Chung2012,Cazalilla2012} through the relation to the entanglement that GGE can be served  as a general ensemble for the integrable model after  a quantum quench. (for a review see \cite{Cazalilla2016}). 

The same concept can be also applied to the entanglement and entanglement spectra\cite{Collura2014,Essler2014,Wright2014,Nezhadhaghighi2014,Torlai2014,Fagotti2015,Canovi2014,Kaufman2016,Apollaro2016,Kogoj2016}.  For the integrable models, the time evolution of entanglement spectra of a system undergoing a sudden quench arrives a steady state at the infinite time\cite{Chung2013,Chung2016}. Considering the simplest case of topological systems with two sublattices in one dimension, for example, the dimerized chains and 1-D $p$-wave superconductors\cite{Kitaev2001}, the Hamiltonian of momentum space has the form of $\mathbf{R}(k)\cdot\boldsymbol{\sigma}$. 
The Berry phase of the system is decided by  $\mathbf{R}(k)$. It  is  $\pi$ symbolizing a topological phase if $\mathbf{R}(k)$ encircles the origin through the whole Brillouin zone, otherwise it is a trivial phase with Berry phase zero\cite{Ryu2002}.   
We are interested in the entanglement measurement for the topological systems undergoing a   sudden quench, i.e.,  $\mathbf{R}$  is suddenly tuned to $\mathbf{R}^{\prime}$ at certain time, and see if the entanglement spectrum reached a steady state or not, moreover, how the edge states evolve. The whole system considered in the  paper is an infinite long chain with periodic boundary conditions. The subsystem  $A$  is a cut in the middle of the chain with length $L$. And the rest of  the chain serves as an environment.
The reduced density matrix $\rho_A$ for a quadratic model can be calculated by the block correlation function matrix, 
$\rho_A = \bigotimes_m
\left[\begin{matrix} \lambda_m & 0\\ 0 & 1-\lambda_m \end{matrix}
 \right]$,
where $\lambda_m$ are the eigenvalues of the correlation
function matrix $G_{i,j} (t)= \tr \rho_0(t) {\hat{\mathbf{c}}}_i \hat{{\mathbf{c}}}_j^{+} $ 
with $\hat{{\mathbf{c}}}_i \equiv (c_i, c_i^{+})^T $ and
$i,j$ being sites of the subsystem $A$ and $\rho_0(t)$ as time evolution of the initial density matrix\cite{Cheong2004,Barthel2006,Peschel2003}.  $\lambda_m$ is know as
the one-particle entanglement spectum (OPES) \footnote{
There are eventually two definitions of one-particle entanglement spectra: The first one is mostly  defined by  Peschel and his coworkers\cite{Peschel2009}  as the one-particle eigenvalues $\varepsilon_m$ of the entanglement Hamiltonian $\mathcal{H}_E$ such as 
\[ \mathcal{H}_E =  \sum_m \varepsilon_m f_m^{\dagger} f_m \] with the fermionic creating and annihilation operators $f_m^{\dagger}$ and $ f_m$,respectively.  (Peschel and his coworkers never mentioned the name ``one-particle entanglement spectra" at all. They just call them eigenvalues of the entanglement Hamiltonian $\mathcal{H}_E$.) The second one is defined by the authors as the one-particle eigenvalues $\lambda_m$ of the reduced density matrix $\rho_A$.  Eventually they are one-to-one correspondent: $\varepsilon_m = \log{\frac{1-\lambda_m}{\lambda_m}}$. The reason why we use the second definition is that by using $\varepsilon_m$ we very quickly encounter number overflows, that means, if $\lambda_m \rightarrow 0 (1) $, $\varepsilon_m \rightarrow \infty (-\infty)$. It is very inconvenient to use $\varepsilon_m$ numerically, therefore we tend to use the second definition. However, they are equivalent.}.
In our previous studies\cite{Chung2013,Chung2016}, we show that the entanglement spectra of such systems reach a steady solution which is described by the effective Hamiltonian $\mathbf{S}_{\mbox{eff}}(k) \cdot\boldsymbol{\sigma}$, where $\mathbf{S}_{\mbox{eff}}=(\hat{\mathbf{R}}\cdot\hat{\mathbf{R}}^\prime) \hat{\mathbf{R}}^{\prime}$ with the unit vectors $\hat{\mathbf{R}}$  and $\hat{\mathbf{R}}^\prime$. Whether  the steady state at  infinite time is topological or not depends on  the  pseudomagnetic field $\mathbf{S}_{\mbox{eff}}$. Therefore the only possibility to obtain a topological phase at  infinite time is  the quench within the same topological phase. For the $p$-wave superconductors we need to put on a further constraint: the gaps of the initial and the final Hamiltonian should not be too far apart to have  a probability that the edge states revive through the dephasing process.

However, the dynamics of the edge states remain smysterious.
We already show in the previous study that the OPES of two edge states remain $1/2$ for a while and then break up in a coherent time proportional to the subsystem size, and relax back to $1/2$ at infinite time. However, experimentally it is very difficult to obtain the infinite time behavior. We still need to know how the system relaxes to their steady states. It can help either the theorists or experimentalists to fit their intermediate time-dependent data. In this paper, we use the saddle point expansion to analyse the power-law decay of the OPES for the edges for dimerized chains and 1-D $p$-wave superconductors.
Those characteristic decay exponents of power-law decay are model dependent. For dimerized chains they are always $3/2$ which is independent of the quench process, while for 1-D $p$-wave superconductors, they can take different values depending on the quench parameters. The lowest value is $1/2$ which leads to a very slow convergence to reach the steady state.  In this way an time-dependent state can be identify as an intermediate state relaxing to a topological state through the decay exponent of the edge.  

The paper is organized as follows: In section~\ref{sec.MM} we  introduce dimerized chains and 1-D $p$-wave superconductors as our starting models. In section~\ref{sec.TEOES} the method to obtain the time evolution of entanglement spectra by using time-dependent correlation function matrices are shown. In section~\ref{sec.R} we show our main results: the exponents of the pow-law relaxation of entanglement spectra are obtained.  Finally, the conclusion and the outlook of our work are presented in section~\ref{sec.SO}. In the \ref{sec.SDE} the mathematic tools of this work, i.e. the steepest decent expansion, are more detailedly illustrated.

\section{Dimerized Chains and 1-D $p$-Wave Superconductors} \label{sec.MM}

\begin{table}[!h]
\centering
\begin{tabular}{c|c|c}
\hline\hline 
 & Dimerized chains & 1-D $p$-wave superconductors \\ 
\hline $\mathbf{c}^\dagger_n$ & $(c^\dagger_\alpha(n),c^\dagger_\beta(n))$ & $(c^\dagger(n),c(n))$ \\
$\mathbf{c}^\dagger_k$ & $(c^\dagger_\alpha(k),c^\dagger_\beta(k))$ & $(c^\dagger(k),c(-k))$ \\
$N_s$ & $2N$ & $N$ \\
$BZ$ & $(-\pi/2,\pi/2]$ & $(-\pi,\pi]$ \\
$V_{BZ}$ & $\pi$ & $2\pi$ \\
$\mathbf{R}(k)$ & $(\delta_-+\delta_+\cos{2k},\delta_+\sin{2k},0)$ & $(0,-\Delta\sin{k},\cos{k}+\mu/2)$ \\ 
\hline\hline
\end{tabular}
\caption{Definitions of the real and momentum space spinors $\mathbf{c}^\dagger_n$ and $\mathbf{c}^\dagger_k$, the number of sites $N_s$, the first Brillouin zone $BZ$ as well as its volume $V_{BZ}$, and the pseudo magnetic field $\mathbf{R}(k)$ in Eq.\eqref{eq.H_k} for dimerized chains and 1-D $p$-wave superconductors. To simplify the notation, we have used $\delta_\pm= 1\pm\delta$.}
\label{tab.def}
\end{table}

In this paper we consider two quadratic fermion systems, which are dimerized chains and 1-D $p$-wave superconductors respectively. The main goal is to investigate how the one-particle entanglement spectrum (OPES) converges to the result of the steady state after a sudden quench. 

For a  dimerized chain, the Hamiltonian reads
\begin{equation}\label{eq.H_DC}
\mathcal{H} = -\sum_{n=1,3,5,\cdots}^{2N-1}{[ (1-\delta)c^\dagger_\beta(n)c_\alpha(n)+\mbox{h.c.} +(1+\delta)c^\dagger_\beta(n)c_\alpha(n+2)+\mbox{h.c.} ]},
\end{equation}
where $N$ is the number of unit cells containing two sites and $\delta\in [-1,1]$ is the the dimerized strength. $c^\dagger_\alpha(n)\equiv c^\dagger(n)$ and $c^\dagger_\beta(n)\equiv c^\dagger(n+1)$ are the fermioic creation operators at odd site $n$ and even site $n+1$ respectively.

On the other hand, the Hamiltonian of the 1-D $p$-wave superconductor is given by
\begin{equation}\label{eq.H_SC}
\mathcal{H} =\sum_{n=1}^{N}{\left[-c^\dagger(n)c(n+1)+\mbox{h.c.}
+\Delta c(n) c(n+1)+\mbox{h.c.}-\mu(c^\dagger(n) c(n) +1/2) \right]},
\end{equation}
where $N$ is the number of sites, $\mu$ is the chemical potential, $\Delta$ is the paring strength (or superconducting gap), and $c^\dagger(n)$ is the creation operator at site $n$.

By using the spinor representation $\mathbf{c}^\dagger_n$ in Table~\ref{tab.def}, imposing periodic boundary condition $\mathbf{c}^\dagger_n=\mathbf{c}^\dagger_{n+N_s}$, and taking the Fourier transformation $\mathbf{c}^\dagger_n=\frac{1}{\sqrt{N}}\sum_{k\in BZ}{e^{-ikn}\mathbf{c}^\dagger_k}$, both Hamiltonians can be re-written as
\begin{equation}\label{eq.H_k}
\mathcal{H}(\mathbf{R}) = -\sum_{k\in BZ}{\mathbf{c}^\dagger_k \left[ \mathbf{R}(k)\cdot\boldsymbol{\sigma} \right]\mathbf{c}_k},
\end{equation}
where $\mathbf{R}(k)\equiv(R_x(k),R_y(k),R_z(k))$ is the real pseudo magnetic field and  $\boldsymbol{\sigma}\equiv(\sigma_x,\sigma_y,\sigma_z)$ is the vector of Pauli matrices. The specific range of the first Brillouin zone and the form of $\mathbf{R}(k)$ are given in Table~\ref{tab.def}. 

\begin{figure}
\center
\includegraphics[width=12.0cm]{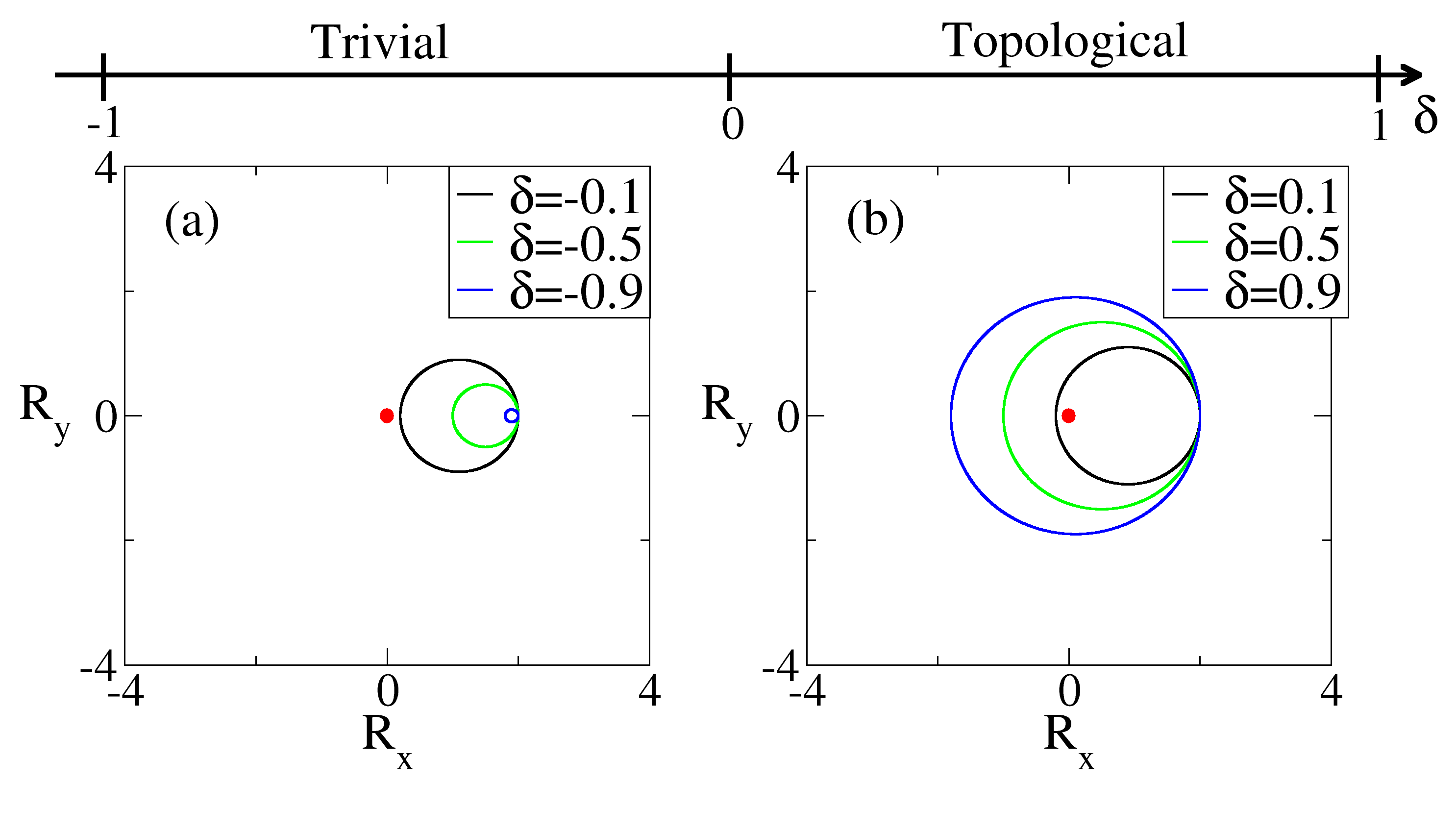}
\caption{(color online) 
The phase diagram of Eq.\eqref{eq.H_DC} and the closed loops $\ell_{\mathbf{R}}$ (circles). The red dot labels the origin. (a) The $\ell_{\mathbf{R}}$ will not enclose the origin in the topological trivial phase. (b) The $\ell_{\mathbf{R}}$ will enclose the origin in the topological phase.}
\label{fig.phase_DC}
\end{figure}

\begin{figure}
\center
\includegraphics[width=12.0cm]{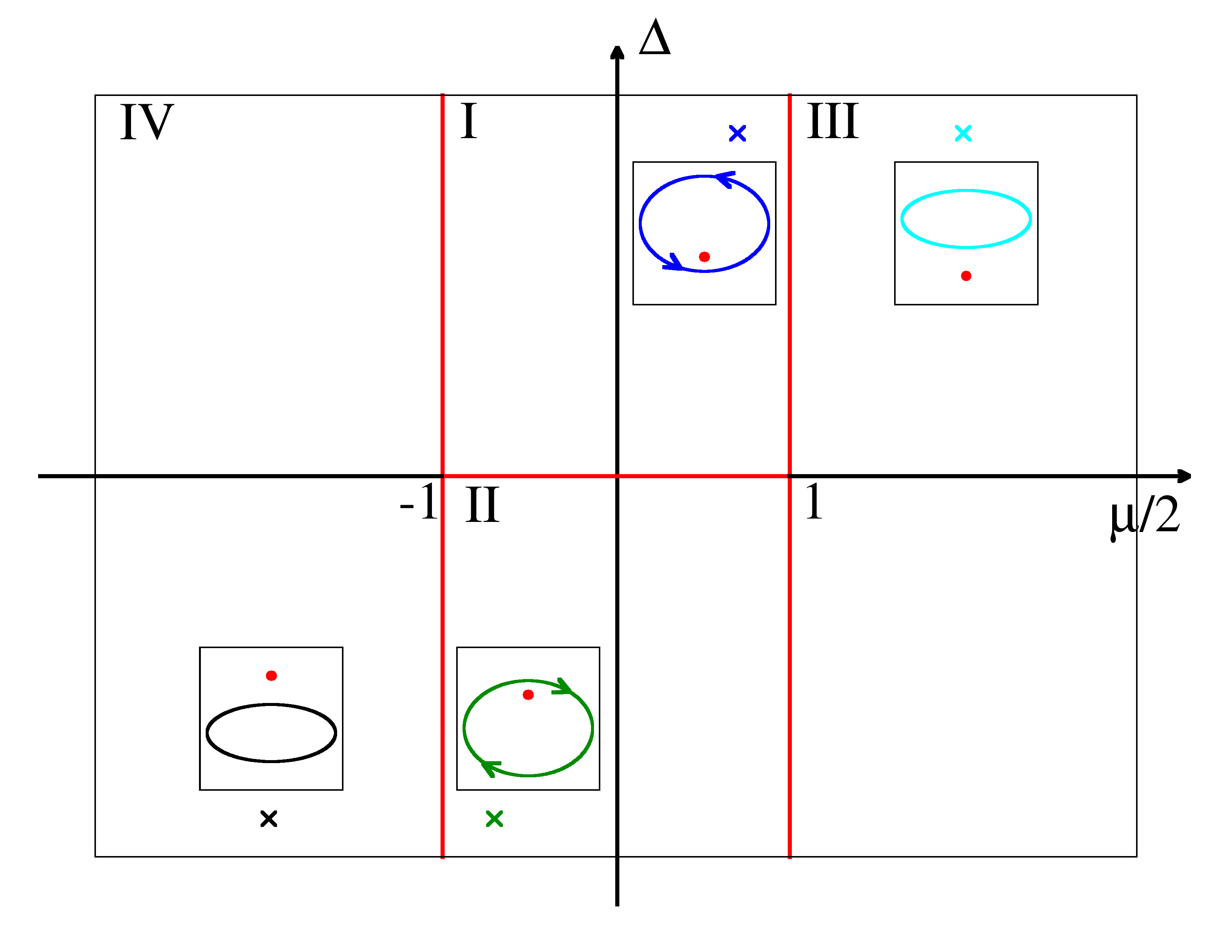}
\caption{(color online) 
The phase diagram of Eq.\eqref{eq.H_SC}. The red lines are the phase boundaries. Each inset shows the closed loop $\ell_{\mathbf{R}}$ using the parameters $(\mu/2,\Delta)$ indicated at the cross point with the same color and the red dot is the origin. Only in the topological phases I and II, the closed loop $\ell_{\mathbf{R}}$ encloses the origin (counterclockwise for phase I and clockwise for phase II).}
\label{fig.phase_SC}
\end{figure}

For both models, the  pseudo magnetic field $\mathbf{R}$ is two dimensional which respects the chiral symmetry. In this case the closed loop $\ell_{\mathbf{R}}$ formed by $\mathbf{R}(k)$ as $k$ runs through the Brillouin zone can be used to characterise the topological properties of the system\cite{Ryu2002}. If $\ell_{\mathbf{R}}$ encloses the origin, the Berry phase of the occupied band is $\pm \pi$ and the system is in a topological phase. In contrast if $\ell_{\mathbf{R}}$ does  not enclose the origin, the Berry phase is zero and the system is in a topological trivial phase. By using this geometric picture, one finds that the dimerized chain is in topological trivial phase if $\delta\in [-1,0)$ and topological phase if $\delta\in(0,1]$ as illustrated in Figure~\ref{fig.phase_DC}. On the other hand, $p$-wave superconductor is in a topological phases if $|\mu/2|<1$ and in a topological trivial phases if $|\mu/2|>1$ as shown in Figure~\ref{fig.phase_SC}. Furthermore, there are two distinct topological phases depending on the clockwise or counterclockwise winding of the closed loop $\ell_{\mathbf{R}}$. All these topological phases are known as the symmetry protected topological (SPT) phases. According to the classification of the SPT phases\cite{Schnyder2008},  the topological phase of the dimerized chains and 1-D $p$-wave superconductors both belongs to the class BDI.

\section{Time Evolution of  Entanglement Spectra} \label{sec.TEOES}
In the following we  illustrate how to obtain the one-particle entanglement spectrum (OPES) $\lambda_m(t)$ for one dimensional quadratic fermionic systems after a sudden quench. Consider a pure state $|\Psi_{AB}\rangle$ in an infinite bipartite system $AB$ with a finite subsystem $A$ and an infinite environment $B$. According to Ref.\cite{Chung2001,Peschel2003,Cheong2004}, the OPES $\lambda_m$ between subsystem $A$ and environment $B$ are  the eigenvalues of the correlation matrix $G$ with matrix element $G_{n,m}\equiv\langle\Psi_{AB}|\mathbf{c}_n \mathbf{c}^\dagger_m|\Psi_{AB}\rangle$. Here $n,m$ are unit cell or site indices for the subsystem $A$ and the spinor $\mathbf{c}^\dagger_n$ is defined in Table~\ref{tab.def}. 

Assume that when $t<0$ the system is in the ground state $|\Psi_G\rangle$ of  $H\equiv\mathcal{H}(\mathbf{R})$.  At  $t=0 $ the system is suddenly quenched to a new Hamiltonian $H^\prime\equiv\mathcal{H}(\mathbf{R}^\prime)$. Starting from Eq.\eqref{eq.H_k}, it is straightforward to show that the Hamiltonian $H$ can be easily diagonalized using the eigenbasis $\mathbf{f}_k=U(\mathbf{R})\mathbf{c}_k$ with the unitary transformation: $U(\mathbf{R})=[(R_z-R)\sigma_z+R_x\sigma_x+R_y\sigma_y]/[2R(R-R_z)]^{\frac{1}{2}}$. The dispersion is then given by $\pm R(k)=\pm (R_x^2+R_y^2+R_z^2)^{\frac{1}{2}}$. Similarly for $H^\prime$ one has eigenbasis $\mathbf{f}^\prime_k=U(\mathbf{R}^\prime)\mathbf{c}_k$. Since at any $t>0$, the system is at a pure state $|\Psi(t)\rangle=e^{-iH^\prime t}|\Psi_G\rangle$, the OPES $\lambda_m(t)$ at time $t$ can be obtained by diagonalizing the time-dependent correlation matrix $G(t)$ with matrix element $G_{n,m}(t)\equiv\langle\Psi(t)|\mathbf{c}_n \mathbf{c}^\dagger_m|\Psi(t)\rangle$, where $G(t)|\lambda_m(t)\rangle=\lambda_m(t)|\lambda_m(t)\rangle$. Due to the translational invariance, we can apply the Fourier transformation and get
\begin{equation}\label{eq.G_nm_def}
  G_{n,m}(t)=\dfrac{1}{N}\sum_{k\in BZ}
  {\;e^{ik(n-m)}\;G_k(t)},
\end{equation}
where 
\begin{equation}
  G_k(t)\equiv \langle\Psi_G | e^{iH^\prime t}  \mathbf{c}_k e^{-iH^\prime t} e^{iH^\prime t} \mathbf{c}^\dagger_k e^{-iH^\prime t}|\Psi_G\rangle.
\end{equation}  

Together with
\begin{equation}
  e^{iH^\prime t}  \mathbf{c}_k e^{-iH^\prime t} 
 = U^\dagger(\mathbf{R}^\prime)  e^{iH^\prime t}  \mathbf{f}^\prime_k e^{-iH^\prime t} 
 = U^\dagger(\mathbf{R}^\prime) e^{-i\epsilon^\prime t \sigma_z}\mathbf{f}^\prime_k,
\end{equation}
where $\epsilon'=R'$ for dimerized chains and $\epsilon'=2R'$ for 1-D $p$-wave superconductors, and
\begin{equation}
  \langle \Psi_ G| \mathbf{f}^\prime_k \mathbf{f}^{\prime \dagger}_k | \Psi_G\rangle
  =   \langle \Psi_ G|  ( U(\mathbf{R}^\prime) U^\dagger(\mathbf{R}) \mathbf{f}_k) ( U(\mathbf{R}^\prime) U^\dagger(\mathbf{R})\mathbf{f}_k)^\dagger  | \Psi_G\rangle,
\end{equation}
it is straightforward to show that
\begin{equation}
  G_k(t)=\dfrac{I}{2}-\dfrac{ (\mathbf{S}_{\mbox{eff}}(k) + \Delta \mathbf{S}(k,t) ) \cdot\boldsymbol{\sigma}}{2},
\end{equation}
where
\begin{equation}\label{eq.S_eff}
  \mathbf{S}_{\mbox{eff}}(k)\equiv(\hat{\mathbf{R}}\cdot\hat{\mathbf{R}}^\prime)\hat{\mathbf{R}}^\prime,
\end{equation}
and
\begin{equation}
  \Delta\mathbf{S}(k, t)\equiv\cos{(2\epsilon^\prime t)}\left[\hat{\mathbf{R}}-(\hat{\mathbf{R}}\cdot\hat{\mathbf{R}}^\prime)
  \hat{\mathbf{R}}^\prime\right]+\sin{(2\epsilon^\prime t)}(\hat{\mathbf{R}}\times\hat{\mathbf{R}}^\prime).
\end{equation}

In the thermodynamic limit $\frac{1}{N}\sum_{k\in BZ}\rightarrow \frac{1}{V_{BZ}}\int_{BZ}{dk}$, and Eq.\eqref{eq.G_nm_def} becomes
\begin{equation}
G_{n,m}(t)
=G_{n,m}^{\mbox{eff}}+\Delta G_{n,m}(t),
\end{equation}
with
\begin{equation}
G_{n,m}^{\mbox{eff}}\equiv \delta_{n,m} \frac{I}{2}-\frac{1}{2V_{BZ}}\int_{BZ}{e^{ik(n-m)}\mathbf{S}_{\mbox{eff}}(k)\cdot\boldsymbol{\sigma}dk}, 
\end{equation}
and
\begin{equation}\label{eq.Delta_G}
\Delta G_{n,m}(t)\equiv-\frac{1}{2V_{BZ}}\int_{BZ}{e^{ik(n-m)}\Delta\mathbf{S}(k,t)\cdot\boldsymbol{\sigma}dk},
\end{equation}
where $V_{BZ}$ is the volume of the first Brillouin zone defined in Table~\ref{tab.def}. According to the Riemann-Lebesgue lemma, one expects that $\Delta G_{n,m}(t) \rightarrow 0$ as $t\rightarrow \infty$, provided that the integrands are smooth functions of $k$. The steady state OPES $\lambda_m(\infty)$ hence can be obtained by diagonalizing $G^{\mbox{eff}}$.

\section{Power-Law Relaxation of Entanglement Spectra} \label{sec.R}

In our previous works\cite{Chung2013,Chung2016}, we have discussed the properties of the effective time-independent pseudo magnetic field $\mathbf{S}_{\mbox{eff}}(k)$ and how it affects the topology of steady state after quench. In this work, however, we are interested in how the OPES $\lambda_m(t)$ approaches to its steady state value $\lambda_m(\infty)$. 
In order to achieve this goal, we first rewrite $\Delta G_{n,m}$ as $\Delta G_{n,m}(t) = \boldsymbol{\mathcal{I}}(t)\cdot\boldsymbol{\sigma}$ with $\boldsymbol{\mathcal{I}}=(\mathcal{I}^x(t),\mathcal{I}^y(t),\mathcal{I}^z(t))$. Furthermore, $\mathcal{I}^x(t),\mathcal{I}^y(t)$, and $\mathcal{I}^z(t)$ can all be written in the form of $\mathcal{I}_R(t)$ in Eq.\eqref{eqA_IR}. 
Using the steepest descent expansion, it is shown in the \ref{sec.SDE} that $\mathcal{I}_R(t)$ behaves asymptotically as a power-law decay superimposed with an oscillating function associated with some characteristic angular frequencies. Consequently, $\Delta G_{n,m}$ also behaves similarly.
Due to its power-law decay, $\Delta G$ can be treated as perturbation at large time, resulting in $\lambda_m(t) \approx \lambda_m(\infty) + \langle \lambda_m (\infty) | \Delta G (t) | \lambda_m (\infty) \rangle$. Because the $|\lambda_m(\infty)\rangle$ is time independent, asymptotically, one has
\begin{equation}\label{eq.goal}
  \lambda_m(t) \sim \lambda_m(\infty) + \frac{Q(t)}{t^\nu},
\end{equation}
where $\nu$ is the smallest power-law decay exponent among all $\mathcal{I}_R(t)$ integrals and $Q(t)$ is an oscillating function that contains all characteristic angular frequencies associated with exponent $\nu$. The details for obtaining the power-law decay exponent $\nu$ and the characteristic angular frequencies are  presented later in this section.

In this section we present our results of the asymptotic behavior of the OPES $\lambda_m(t)$ (Eq.\eqref{eq.goal})   for dimerized chains and 1-D $p$-wave superconductors by performing the steepest descent expansion to Eq.\eqref{eq.Delta_G}.  In  order to have a non-trivial dynamics, we  assume that the initial and final Hamiltonians do not commute with each other and furthermore the final Hamiltonian does not have a constant dispersion. Otherwise, the time evolving state $|\Psi(t)\rangle$  remain the same up to a global phase difference.
To confirm our theoretical results, we can  numerically find the time evolution of the one-particle entanglement spectra $\lambda(t)$ closest to $1/2$. Due to the particle-hole symmetry, $\lambda(t)$ and its particle-hole pair $1-\lambda(t)$  behave in the same way. Hence, we only show  $\lambda(t)$ larger than $1/2$. 

\subsection{Dimerized chains}

\begin{figure}
\center
\includegraphics[width=10cm]{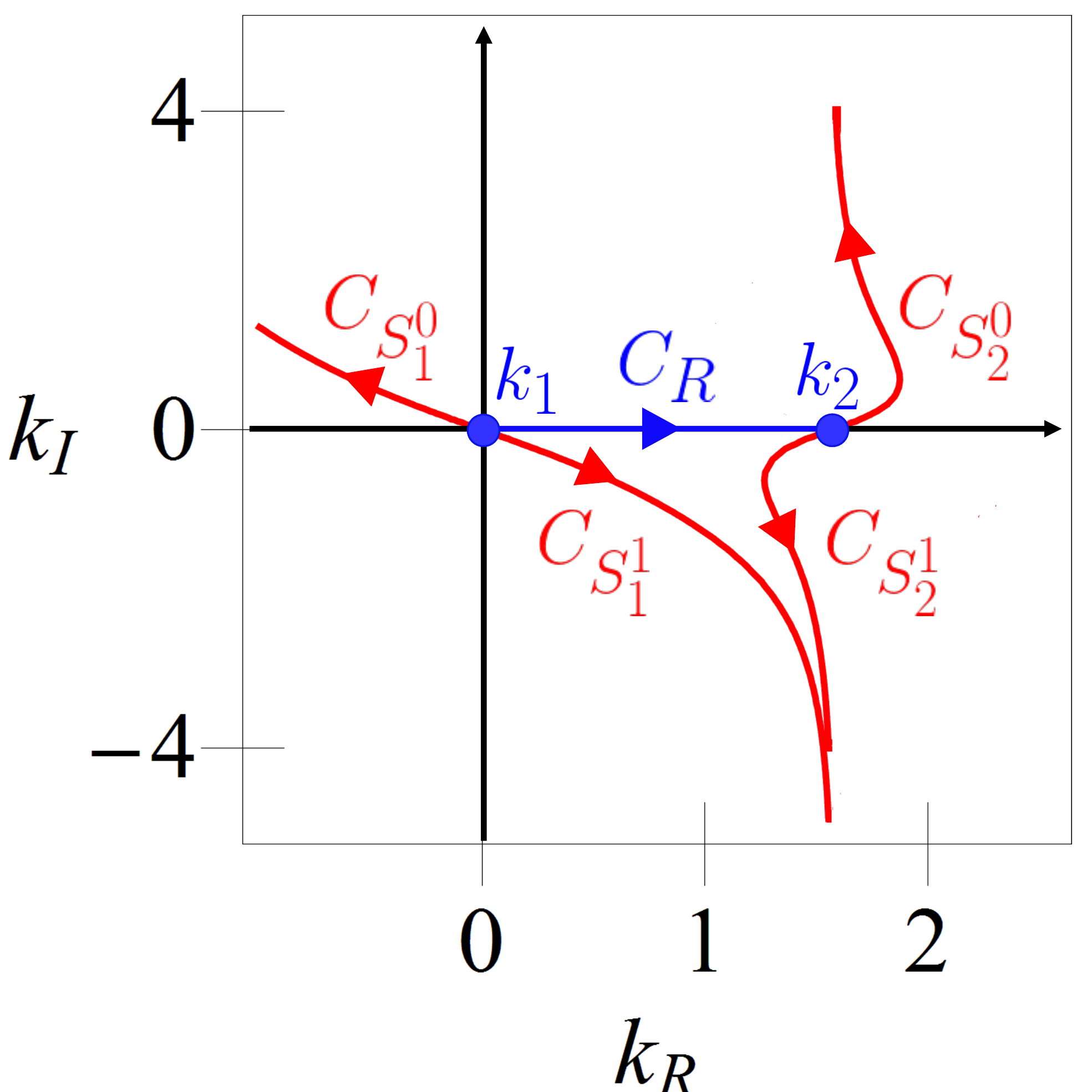}
\caption{(color online) 
The typical steepest descent paths $C_{S^l_l}$ (red curves) starting from a saddle point $k_j$ (blue dots) and the original integration path $C_{R}$ (blue line). The path $C_R-C_S$ forms a closed loop with $C_S=C_{S^1_1}-C_{S^1_2}$. The parameter we used here is $\delta'=-0.4$.}
\label{fig.path_DC}
\end{figure}

\begin{figure}
\center
\includegraphics[width=13.0cm]{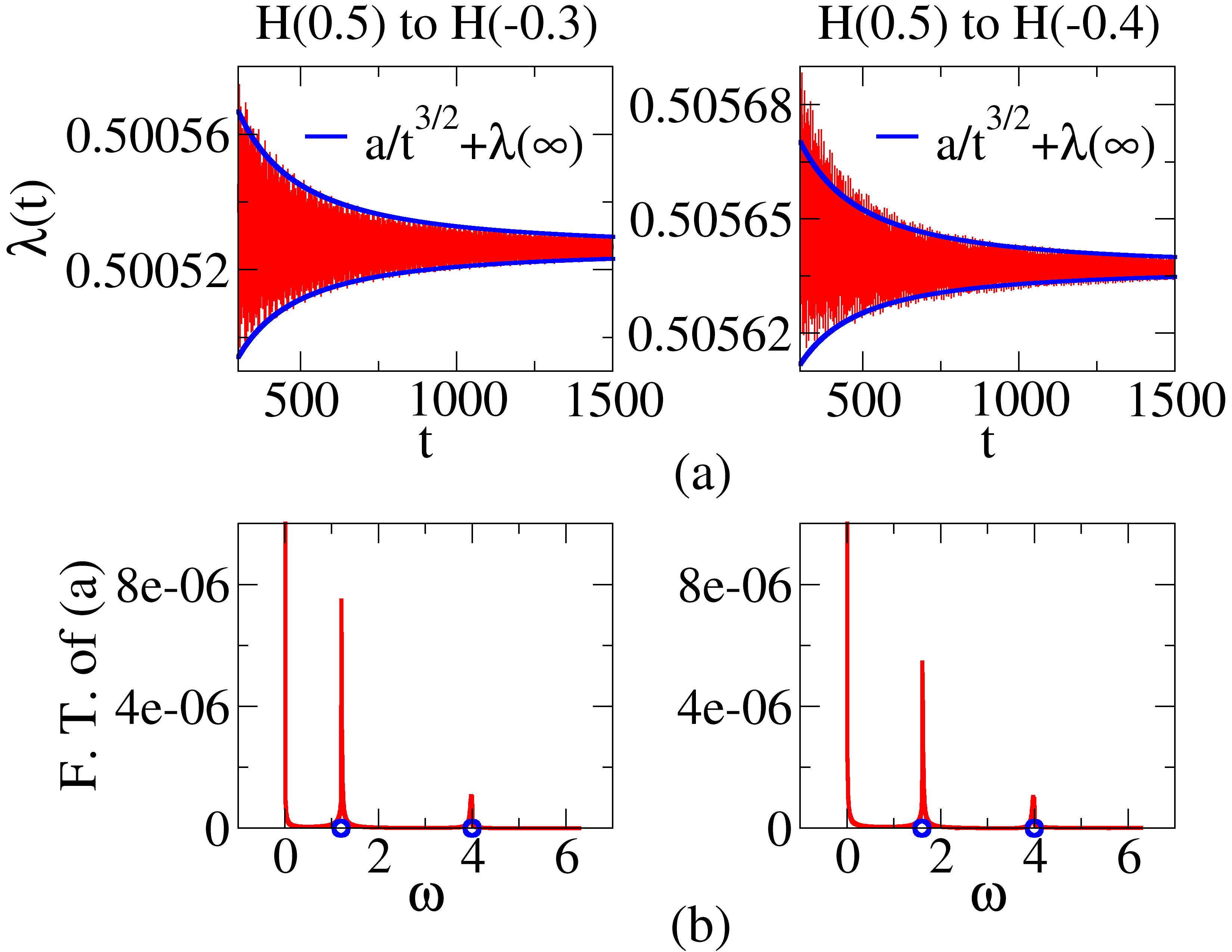}
\caption{(color online) The numerical results for the quench from $\delta=0.5$ to $\delta'=-0.3$ (left) and $\delta'=-0.4$ (right) with subsystem size $L=200$. (a) The most important OPES $\lambda(t)$ (red) and the power-law fitting $at^{-3/2}+\lambda(\infty)$ (blue) to the envelopes of $\lambda(t)$. (b) The Fourier transformation of (a) for large time region. The two  characteristic angular frequencies (open blue circles) at $4$ and $4|\delta^\prime|$ ($1.2$ for the left and $1.6$ for the right case) are found. Note that the peak at $\omega=0$ is due to the steady OPES $\lambda(\infty)$}
\label{fig.result_DC}
\end{figure}

\begin{figure}
\center
\includegraphics[width=13.0cm]{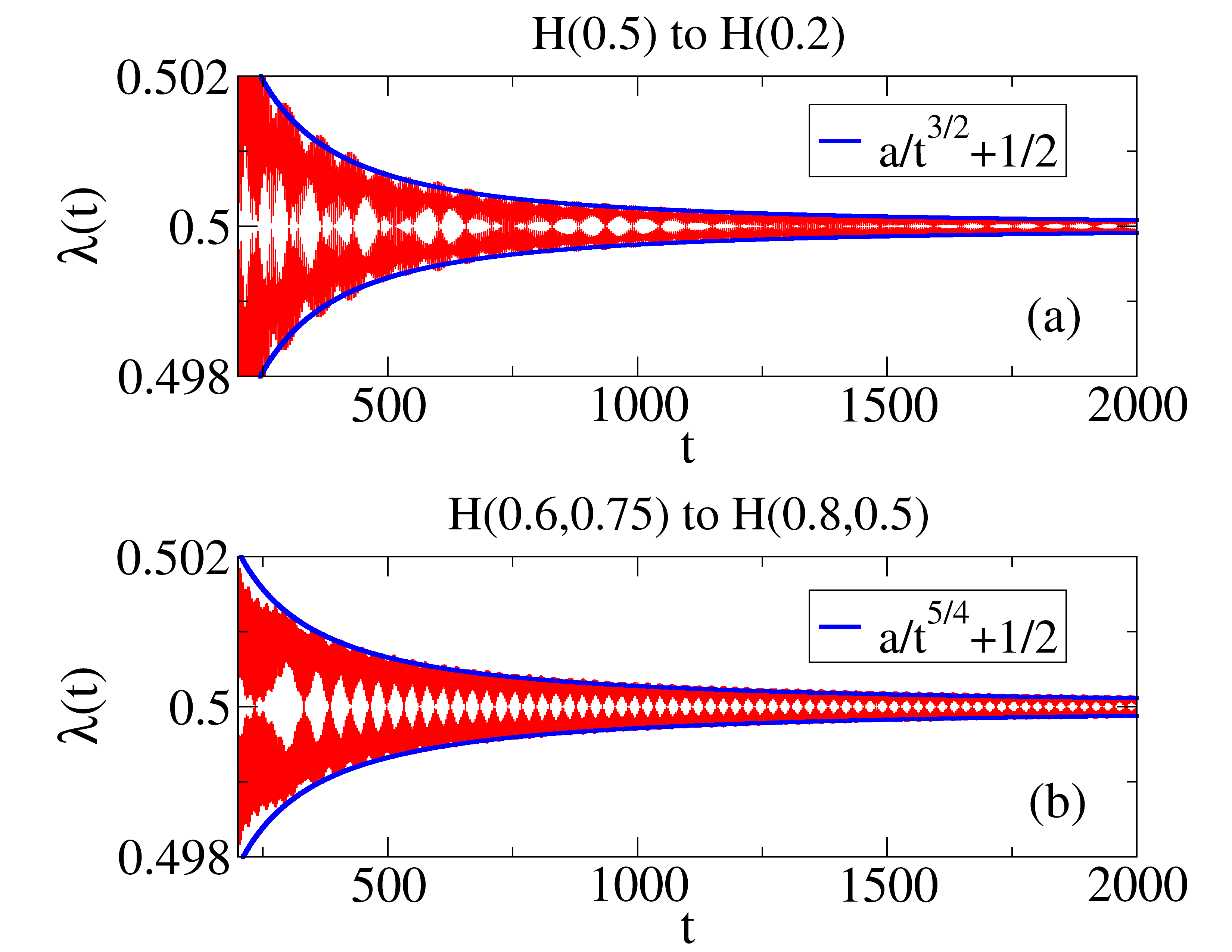}
\caption{
(color online) The numerical results in the case that the subsystem ends up with a topological steady state after quench. Results for (a) dimerized chains ($\delta=0.5$ to $\delta'=0.3$) with subsystem size $L=200$ and (b) 1-D $p$-wave superconductors in class II ($\mu/2,\Delta=0.6,0.75$ to $\mu'/2=0.8,0.5,\delta'=0.4$) with subsystem size $L=100$. As one can see, the most important OPES $\lambda(t)$ (red) fit the predicted power-law decay (blue) very well.
}
\label{fig.result_steady_topological}
\end{figure}

In this subsection we consider a sudden quench of a dimerized chain from $H=\mathcal{H}(\delta)$ to $H^\prime=\mathcal{H}(\delta^\prime)$. In order to study how the OPES $\lambda_m(t)$ approach to their steady state values, we first rewrite Eq.\eqref{eq.Delta_G} as $\Delta G_{n,m}(t) = \boldsymbol{\mathcal{I}}(t)\cdot\boldsymbol{\sigma}$ and find

\begin{equation}
  \mathcal{I}^x(t) =  \int_0^{\pi/2}{\cos{[k(n-m)]}\left\{\hat{R}_x(k)-\left[\hat{\mathbf{R}}(k)\cdot\hat{\mathbf{R}}'(k)\right]\hat{R}_x'(k)\right\}\cos{\left[ 2 R^\prime(k) t\right]}dk}, 
\end{equation}

\begin{equation}
  \mathcal{I}^y(t) = i\int_0^{\pi/2}{\sin{[k(n-m)]}\left\{\hat{R}_y(k)-\left[\hat{\mathbf{R}}(k)\cdot\hat{\mathbf{R}}'(k)\right]\hat{R}_y'(k)\right\}\cos{\left[2 R^\prime(k) t\right]}dk},
\end{equation}

\begin{equation}
  \mathcal{I}^z(t) = i\int_0^{\pi/2}{\sin{[k(n-m)]}\left[\hat{R}_x(k)\hat{R}_y'(k)-\hat{R}_y(k)\hat{R}_x'(k)\right]\sin{\left[2 R^\prime(k) t\right]}dk},
\end{equation}
where $R=|\mathbf{R}|$, $\hat{\mathbf{R}}\equiv \mathbf{R}/R$ and $\hat{R}_{x,y,z}\equiv R_{x,y,z}/R$ . During the derivation we have also used the properties that $R_x(k)=R_x(-k), R_y(k)=-R_y(-k)$ and $R_z(k)=0$.
The asymptotic behavior of $\boldsymbol{\mathcal{I}}(t)$ as $t\rightarrow \infty$ can be obtained by applying the steepest decent expansion.
To proceed we rewrite $\mathcal{I}^x(t)$ as 
\begin{equation}
  \mathcal{I}^x(t) =  \text{Re} \left[
    \int_{C_R}  f(k) e^{i2 R^\prime(k) t}dk
  \right]
\end{equation}
where $f(k)=\cos{[k(n-m)]}[\hat{R}_x-(\hat{\mathbf{R}}\cdot\hat{\mathbf{R}}')\hat{R}_x']$ and $C_R=[0,\pi/2]$. For $\mathcal{I}^y(t)$ and $\mathcal{I}^z(t)$ we proceed in a similar fashion. By using the steepest descent method as outlined in \ref{sec.SDE}, we find there are two order-2 saddle points, therefore $n_1=n_2=2$, located at $k_1=0$ and $k_2=\pi/2$ respectively. The steepest descent path $C_S$ can be chosen in the way indicated in Figure~\ref{fig.path_DC}. The power-law decay exponents $\nu_j$ is defined as $\beta_j/n_j$, where $\beta_j-1$ is the leading order of the Taylor expansion of $f(k)$ around the saddle point (see Eq.\eqref{eqA_f}). As a result,  $\beta^{x,y,z}_1=\beta^{x,y,z}_2=3$ and $\nu^{x,y,z}_1=\nu^{x,y,z}_2=3/2$. The OPES asymptotically behaves as 
\begin{equation}\label{eq.result_DC}
  \lambda_m(t) \sim \lambda_m(\infty) + \frac{Q(t)}{t^{3/2}},
\end{equation}
where $Q(t)$ contains two oscillating angular frequencies:  $\omega_1=|w(k_1)|=4$ and $\omega_2=|w(k_2)|=4|\delta^\prime|$(see Eq.\eqref{eqA_omega}). 
In Figure~\ref{fig.result_DC}(a) we show our numerical results of the most important OPES $\lambda(t)$ which is larger and closest to $1/2$ for two different $\delta^\prime$. It is clear that $at^{-3/2}+\lambda(\infty)$ can fit the envelopes of $\lambda(t)$ well in both cases. In Figure~\ref{fig.result_DC}(b) we show the Fourier transform of $\lambda(t)$ and two peaks around $\omega_1$ and $\omega_2$ are clearly observed. This confirms that the characteristic frequencies of $Q(t)$ are indeed $\omega_1$ and $\omega_2$. If the subsystem ends up with a topological steady sate, i.e. both the two most important OPES approach to $1/2$ and become degenerate as time goes to infinity. Although finding the characteristic frequencies becomes numerically difficult due to the oscillation around the steady value $1/2$, one can still obtain a nice fit of the envelopes to our predicted power-law decay with exponent $3/2$ as shown in Figure~\ref{fig.result_steady_topological}(a).

\subsection{1-D $p$-wave superconductors}

\begin{figure}[!h]
  \centering 
  \includegraphics[width=10cm]{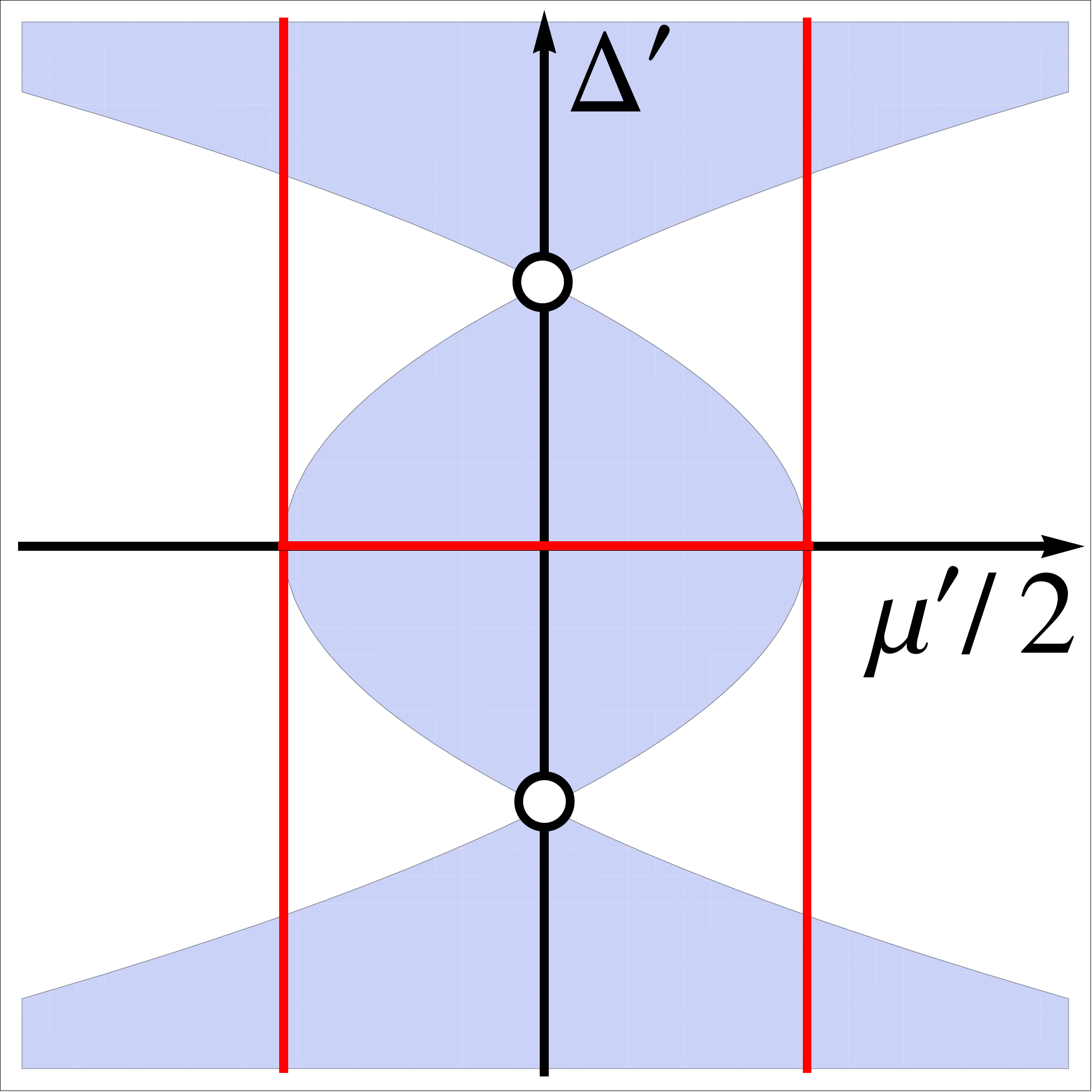}
  \caption
  {(color online)
    The regimes for class I (blue), II (boundary between blue and white) and III (white) in the phase diagram. The red lines are phase boundaries. Note that the two open circles at $(\mu'/2,\Delta')=(0,\pm 1)$ are not under our consideration due to the constant dispersion, which gives a trivial dynamics. 
  }
\label{fig.class_SC}
\end{figure}

We next consider a sudden quench of a 1-D $p$-wave superconductors from $H=\mathcal{H}(\mu/2,\Delta)$ to  $H^\prime=\mathcal{H}(\mu'/2,\Delta')$. Again, we rewrite Eq.\eqref{eq.Delta_G} as $\Delta G_{n,m}(t) = \boldsymbol{\mathcal{I}}(t)\cdot\boldsymbol{\sigma}$ and find

\begin{equation}
  \mathcal{I}^x(t) = i\int_0^{\pi}{\sin{[k(n-m)]}\left[\hat{R}_y(k)\hat{R}'_z(k)-\hat{R}_z(k)\hat{R}'_y(k)\right]\sin{\left[ 4 R'(k) t\right]}dk}
\end{equation}

\begin{equation}
  \mathcal{I}^y(t) = i\int_0^{\pi}{\sin{[k(n-m)]}\left\{\hat{R}_y(k)-\left[\hat{\mathbf{R}}(k)\cdot\hat{\mathbf{R}}'(k)\right]\hat{R}'_y(k)\right\}\cos{\left[4 R^\prime(k) t\right]}dk}
\end{equation}

\begin{equation}
  \mathcal{I}^z(t) = \int_0^{\pi}{\cos{[k(n-m)]}\left\{\hat{R}_z(k)-\left[\hat{\mathbf{R}}(k)\cdot\hat{\mathbf{R}}'(k)\right]\hat{R}'_z(k)\right\}\cos{\left[ 4 R^\prime(k) t\right]}dk} 
\end{equation}
During the derivation we have also used the properties that $R_z(k)=R_z(-k), R_y(k)=-R_y(-k)$ and $R_x(k)=0$.
The asymptotic behavior of $\boldsymbol{\mathcal{I}}(t)$ as $t\rightarrow \infty$ can be again obtained by applying the steepest decent expansion.
To proceed we rewrite $\mathcal{I}^x(t)$ as 
\begin{equation}
  \mathcal{I}^x(t) =  \text{Re} \left[
    \int_{C_R}  f(k) e^{i4 R^\prime(k) t}dk
  \right]
\end{equation}
where $f(k)=\sin{[k(n-m)]}[\hat{R}_x-(\hat{\mathbf{R}}\cdot\hat{\mathbf{R}}')\hat{R}'_x]$ and $C_R=[0,\pi]$. For $\mathcal{I}^y(t)$ and $\mathcal{I}^z(t)$ we proceed similarly. The steepest descent analysis tells that there are three possible saddle points $k_1=0$, $k_2=\pi$, and $k_3=\cos^{-1}{[(\mu'/2)/({\Delta'}^2 -1)]}$ near $C_R$. However, different from the dimerized chains, the orders of the saddle points depend on the value of $k_3$. 
In order to proceed, we thus classify the situations into class I, II and III as the case $|(\mu'/2)/({\Delta'}^2 -1)|<1$, $|(\mu'/2)/({\Delta'}^2 -1)|=1$, and $|(\mu'/2)/({\Delta'}^2 -1)|>1$, respectively. A pictorial illustration for the three classes in the phase diagram is shown in Figure~\ref{fig.class_SC}.

\subsubsection{Class I: $|\frac{\mu'/2}{{\Delta'}^2 -1}|<1$}

\begin{figure}[!h]
  \centering 
  \includegraphics[width=10cm]{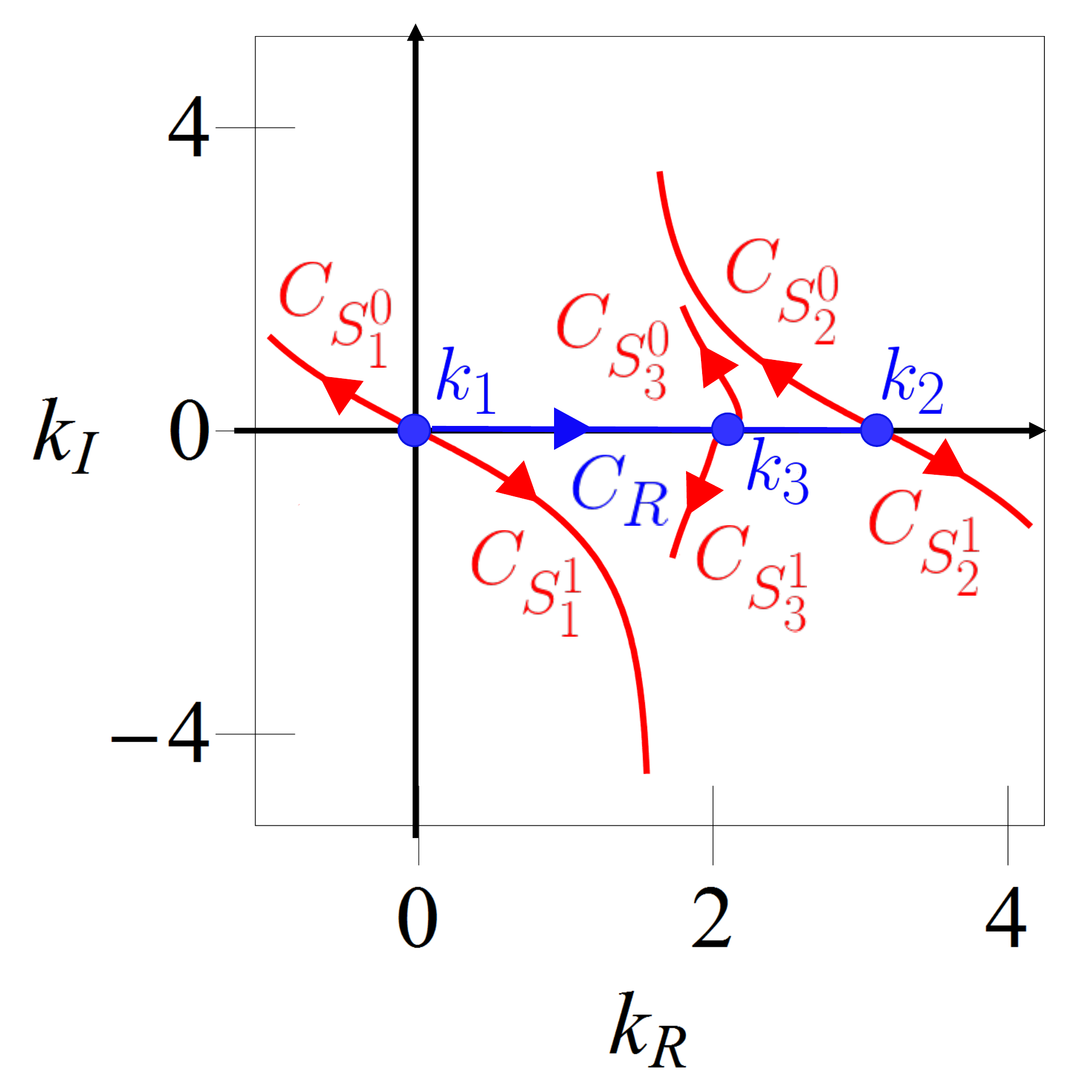}
  \caption
  {(color online)
    The typical steepest descent paths $C_{S^l_l}$ (red curves) started from the saddle points $k_j$ (blue dots) and the original integration path $C_R$ (blue line) in class I. The path $C_R-C_S$ can form a closed loop by choosing $C_S=C_{S^1_1}-C_{S^1_3}+C_{S^0_3}-C_{S^0_2}$. We use $(\mu'/2,\Delta')=(0.5,0.1)$ for this figure.  
  }
\label{fig.path_I_SC}
\end{figure}

\begin{figure}
\center 
\includegraphics[width=12cm]{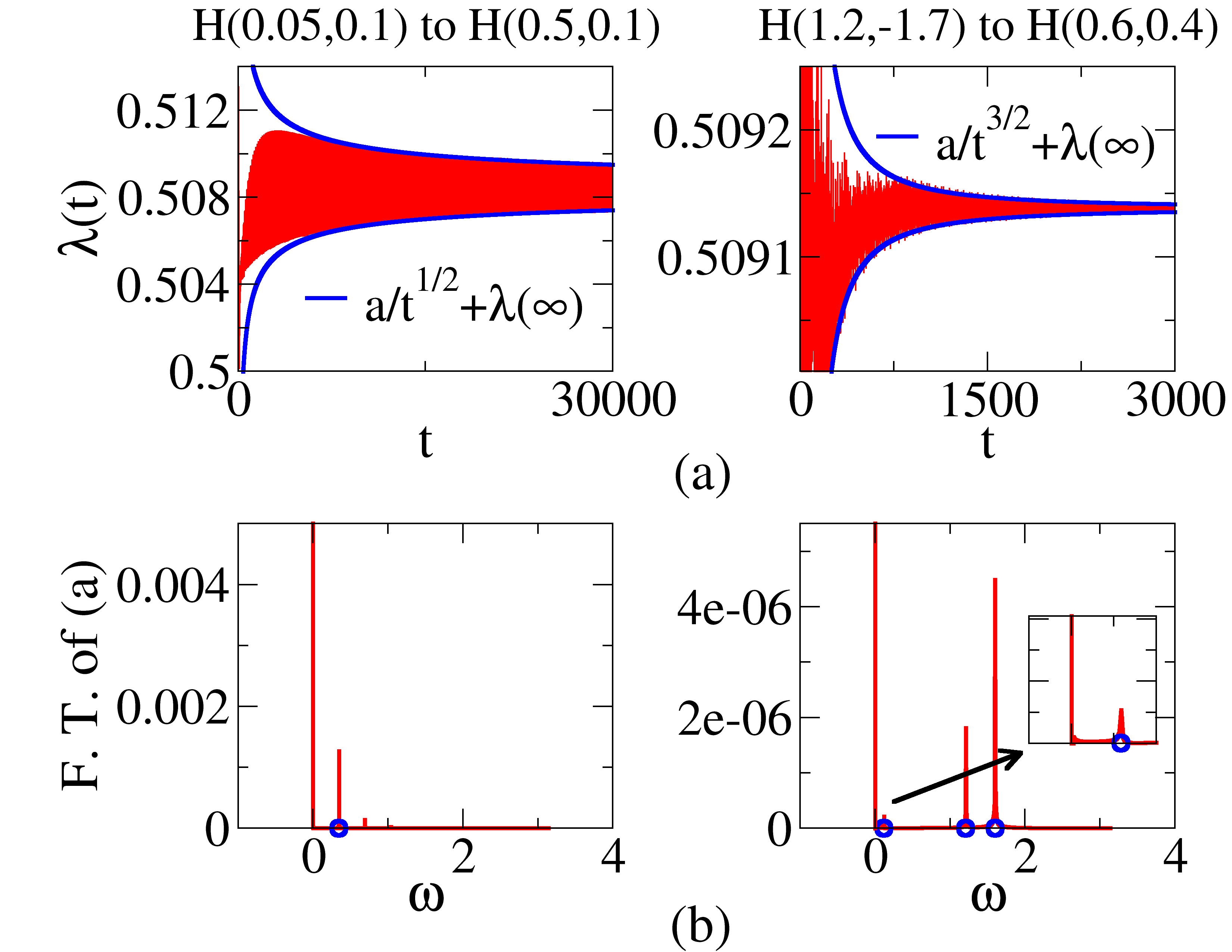}
\caption{(color online) 
The numerical calculations in class I for the cases $\frac{\mu'-\mu}{\Delta'(\mu'\Delta-\mu\Delta')}=100$ (left) and $\frac{\mu'-\mu}{\Delta'(\mu'\Delta-\mu\Delta')}=1$ (right together with the condition $\beta^q_1=\beta^q_2=3$) with subsystem size $L=100$. (a) The most important OPES $\lambda(t)$ (red) and the power-law fittings $at^{-\nu}+\lambda(\infty)$ to the envelopes of $\lambda(t)$, where $\nu=1/2$ for left case and $\nu=3/2$ for the right case. (b) The Fourier transformation of (a). The characteristic angular frequency (open blue circle) $\omega_3\approx 0.345826$ for left panel and $\omega_1, \omega_2, \omega_3\approx 0.117, 1.209, 1.6$ for the right panel are found. The peak at $\omega=0$ comes from the steady OPES $\lambda(\infty)$.}
\label{fig.result_I_SC}
\end{figure}

In class I, $k_1$, $k_2$ and $k_3$ are all order-2 saddle points $n_j=2$ and we can choose the steepest descent paths $C_{S}$ as illustrated in Figure~\ref{fig.path_I_SC}. On the other hand, the $\beta^{x,y,z}_j$, are
\begin{equation}\label{eq.beta1_SC}
\beta^{x,y,z}_1=
\begin{cases}
5 & \mbox{if} \;\;\frac{\Delta'}{\Delta}=\frac{1+\mu'/2}{1+\mu/2} \\
3 & \mbox{if otherwise}
\end{cases},
\end{equation}
\begin{equation}\label{eq.beta2_SC}
\beta^{x,y,z}_2=
\begin{cases}
5 & \mbox{if} \;\;\frac{\Delta'}{\Delta}=\frac{1-\mu'/2}{1-\mu/2}  \\
3 & \mbox{if otherwise}
\end{cases}, 
\end{equation}
\begin{equation}\label{eq.beta3_SC}
\begin{aligned}
&\beta^x_3=\beta^z_3=
\begin{cases}
2 & \mbox{if} \;\;\frac{\mu'-\mu}{\Delta'(\mu'\Delta-\mu\Delta')}=1  \\
1 & \mbox{if otherwise}
\end{cases}, \\
&\beta^y_3=
\begin{cases}
3 & \mbox{if} \;\;\mu'=\mu=0 \\
2 & \mbox{if} \;\;\frac{\mu'-\mu}{\Delta'(\mu'\Delta-\mu\Delta')}=1  \\
1 & \mbox{if otherwise}
\end{cases}.
\end{aligned}
\end{equation}

\begin{table}
\centering
(a) \\
\begin{tabular}{c|c|c|c}
\hline\hline 
 & $x$ & $y$ & $z$ \\ 
\hline 
$j=1$ & $5/2$\;\mbox{or}\;$3/2$ & $5/2$\;\mbox{or}\;$3/2$ & $5/2$\;\mbox{or}\;$3/2$ \\
$j=2$ & $5/2$\;\mbox{or}\;$3/2$ & $5/2$\;\mbox{or}\;$3/2$ & $5/2$\;\mbox{or}\;$3/2$ \\
$j=3$ & $1/2$ & $1/2$\;\mbox{or}\;$3/2$ & $1/2$ \\
\hline\hline
\end{tabular}
\\
(b)
\\
\begin{tabular}{c|c|c|c}
\hline\hline 
 & $x$ & $y$ & $z$ \\ 
\hline 
$j=1$ & $5/2$\;\mbox{or}\;$3/2$ & $5/2$\;\mbox{or}\;$3/2$ & $5/2$\;\mbox{or}\;$3/2$ \\
$j=2$ & $5/2$\;\mbox{or}\;$3/2$ & $5/2$\;\mbox{or}\;$3/2$ & $5/2$\;\mbox{or}\;$3/2$ \\
$j=3$ & $3/2$ & $3/2$ & $3/2$ \\
\hline\hline
\end{tabular}
\caption{Table of the power-law decay exponents $\nu^{x,y,z}_j$ for (a) $\frac{\mu'-\mu}{\Delta'(\mu'\Delta-\mu\Delta')}\neq1$ and (b) $\frac{\mu'-\mu}{\Delta'(\mu'\Delta-\mu\Delta')}=1$. The dominate exponent $\nu^{x,y,z}$ for $\mathcal{I}^{x,y,z}(t)$ is the smallest one for each column. Note that the exponent is defined as  $\beta^{x,y,z}_j/n_j$ except the case of $j=3$ and $\frac{\mu'-\mu}{\Delta'(\mu'\Delta-\mu\Delta')}=1$, where the exponent is defined as $1+1/n_3$ because $\beta^{x,y,z}_3=n_3$.}
\label{tab.nu_q_I_SC}
\end{table}

When $\frac{\mu'-\mu}{\Delta'(\mu'\Delta-\mu\Delta')}\neq1$, the smallest $\nu$ equals $1/2$ as shown in Table~\ref{tab.nu_q_I_SC}(a), and the decay behavior is dominated by the saddle point $k_3$. We hence conclude that the asymptotic behavior of the OPES $\lambda_m(t)$ is a power-law decay with exponent $1/2$ and angular frequency

\begin{equation} \label{omega3}
  \omega_3=|w(k_3)|=4|\Delta'|\sqrt{({\Delta'}^2+{(\mu'/2)}^2-1)/({\Delta'}^2-1)}.
\end{equation}

In the case of $\frac{\mu'-\mu}{\Delta'(\mu'\Delta-\mu\Delta')}=1$, however, one has $\beta^{x,y,z}_3=n_3=2$. As shown in the \ref{sec.SDE}, when $\beta=n$ the leading terms cancel each other and one needs to calculate the next-leading term. Consequently one has $\nu^{x,y,z}_3=1+1/n_3$ instead of $\beta^{x,y,z}_3/n_3$. In this case we find the smallest $\nu$ equals $3/2$ as shown in Table~\ref{tab.nu_q_I_SC}(b).
However, different from the previous case, at least one of the two saddle points $k_1$ and $k_2$ also contribute to this exponent as long as $\beta^{x,y,z}_1=3$ or $\beta^{x,y,z}_2=3$. As a result  $\lambda_m(t)$ asymptotically behaves like  a power-law decay with exponent $\nu=3/2$ with at least one angular frequencies $\omega_3$ Eq.\eqref{omega3}).

We summarize the above two cases as follows
\begin{equation}\label{eq.result_I_SC}
\lambda_m(t) \sim \lambda_m(\infty) +
\begin{cases}
\frac{Q(t)}{t^{3/2}} & \mbox{if} \;\; \frac{\mu'-\mu}{\Delta'(\mu'\Delta-\mu\Delta')}=1 \\
\frac{Q'(t)}{t^{1/2}} & \mbox{if otherwise}
\end{cases},
\end{equation}
where $Q(t)$ contains three possible oscillating angular frequencies: $\omega_1=|w(k_1)|=4|\mu'/2+1|$ (if $\beta^{x,y,z}_1=3$), $\omega_2=|w(k_2)|=4|\mu'/2-1|$ (if $\beta^{x,y,z}_2=3$) as well as $\omega_3$ Eq.\eqref{omega3} and $Q'(t)$ has only one oscillating angular frequency $\omega_3$. As demonstrated in Figure~\ref{fig.result_I_SC}(a), our numerical calculations of the most important OPES $\lambda(t)$ larger and closest to $1/2$ show a good agreement with the power-law fittings to the envelopes of $\lambda(t)$ in both cases of Eq.\eqref{eq.result_I_SC}. Besides, the Fourier transformation of $\lambda(t)$ shown in Figure~\ref{fig.result_I_SC}(b) also confirms the existences of the characteristic angular frequencies we have predicted, where the peaks at the predicted frequencies (blue circles) are found.

\subsubsection{Class II: $|\frac{\mu'/2}{{\Delta'}^2 -1}|=1$}
\begin{figure}[!h]
  \centering 
  \includegraphics[width=10cm]{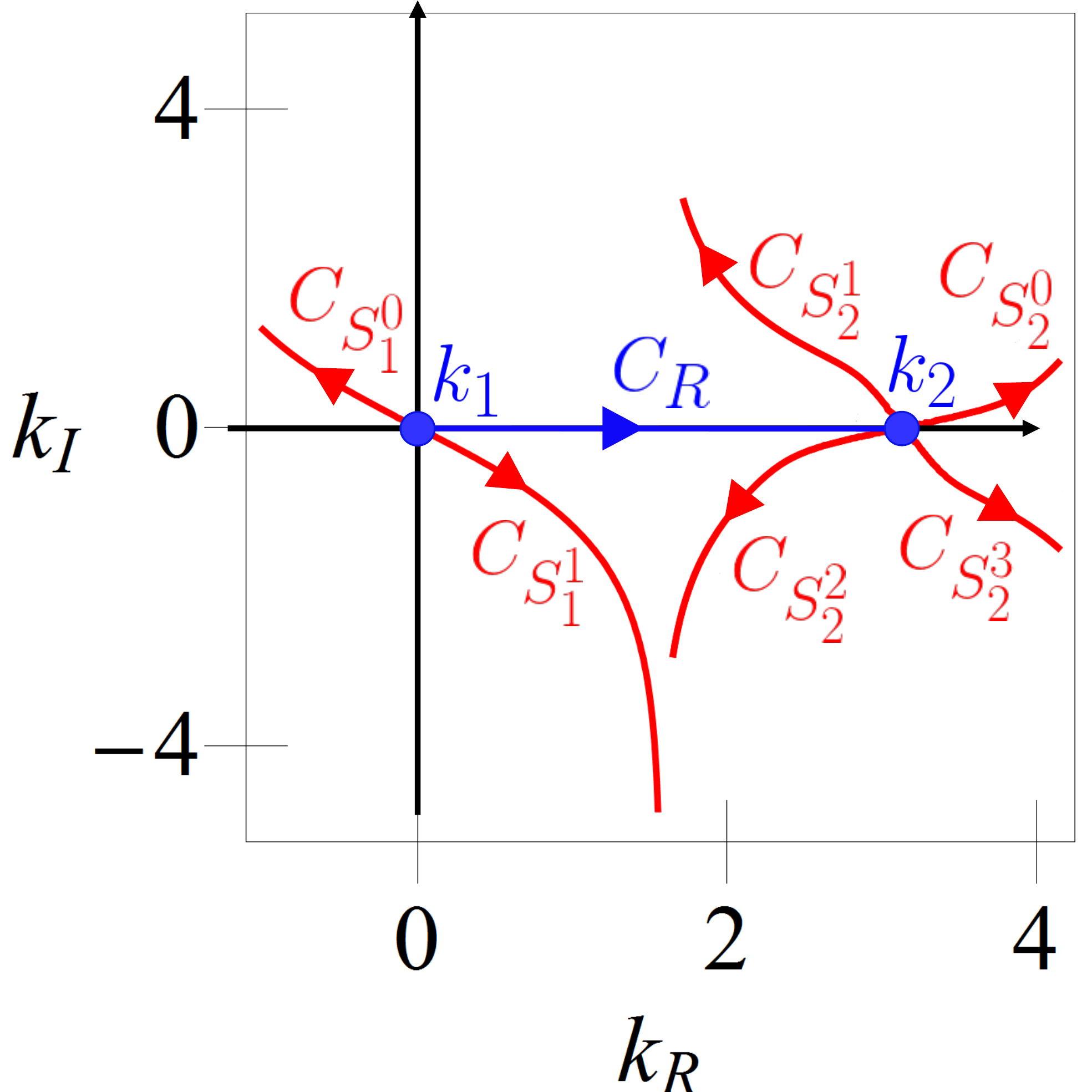}
  \caption
  {(color online)
The typical steepest descent paths $C_{S^l_j}$ (red curves) started from the saddle points $k_j$ (blue dots) and the original integration path $C_R$ (blue line) in class II. One can choose $C_S=C_{S^1_1}-C_{S^2_2}$ such that the path $C_R-C_S$ forms a closed loop. We use $(\mu'/2,\Delta')=(0.75,0.5)$ for this figure. 
  }
\label{fig.path_II_SC}
\end{figure}

\begin{figure}
\center 
\includegraphics[width=12cm]{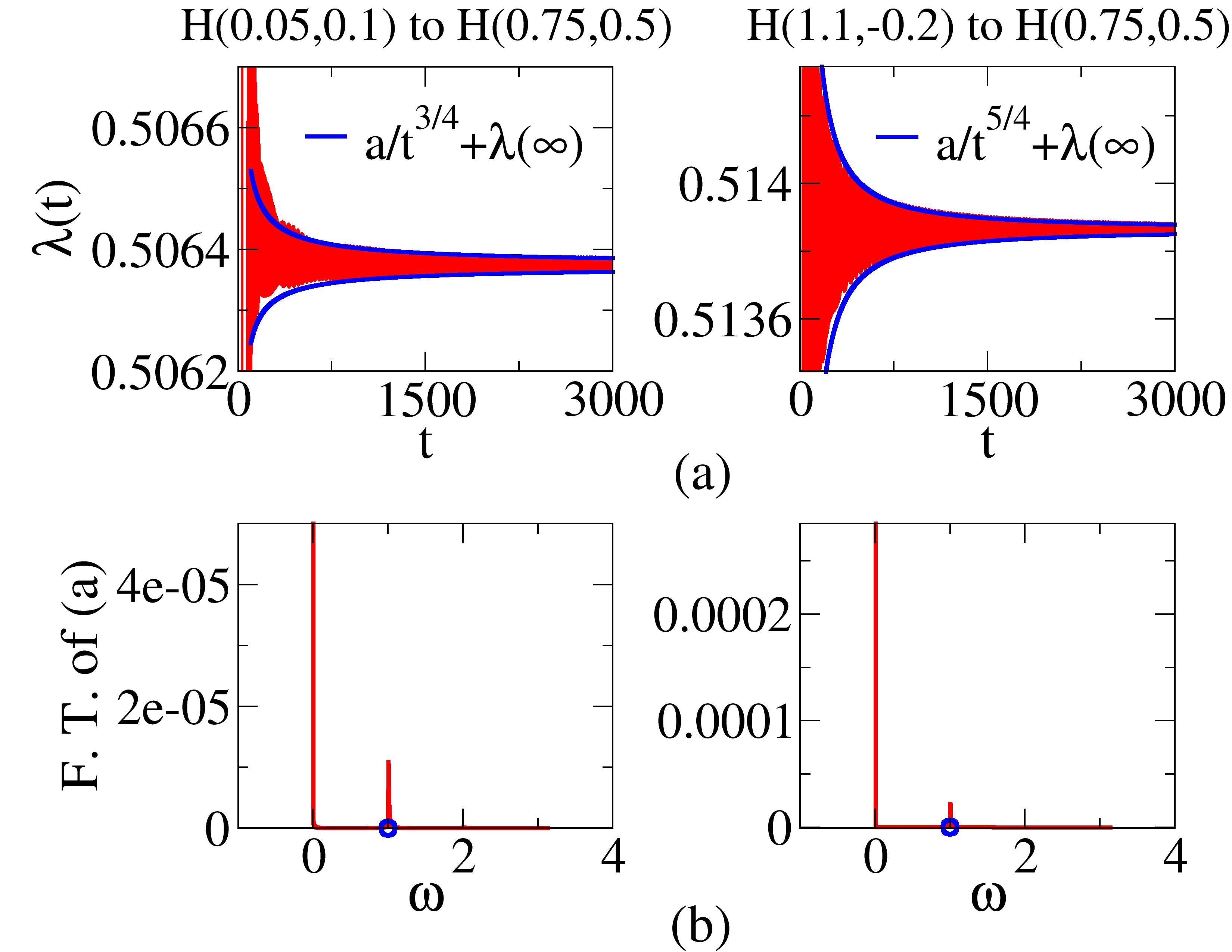}
\caption{(color online) 
The numerical calculations in class II for the case $\Delta'/\Delta\neq(1\pm\mu'/2)/(1\pm\mu/2)$ (left) and the case $\Delta'/\Delta=(1-\mu'/2)/(1-\mu/2)$ (right) with subsystem size $L=100$ and $(\mu'/2)/({\Delta'}^2-1)=-1$. (a) The most important OPES $\lambda(t)$ (red) and the power-law fittings $at^{-\nu}+\lambda(\infty)$ to the envelopes of $\lambda(t)$, where $\nu=3/4$ for left panel and $\nu=5/4$ for the right panel. (b) The Fourier transformation of (a). The characteristic angular frequency (open blue circle) $\omega_2=1$ for the left panel and $\omega_2=1.25$ for the right panel are found. The peak at $\omega=0$ comes  from the steady OPES $\lambda(\infty)$.}
\label{fig.result_II_SC}
\end{figure}

In class II, we have only two saddle points $k_1$ and $k_2$ since $k_3$ coincide with either $k_1$ or $k_2$. As a result, the order of the saddle points are 
\begin{equation}
n_1=
\begin{cases}
4 & \mbox{if} \;\;\frac{\mu'/2}{{\Delta'}^2 -1}=1  \\
2 & \mbox{if} \;\;\frac{\mu'/2}{{\Delta'}^2 -1}=-1
\end{cases},
\end{equation}
\begin{equation}
n_2=
\begin{cases}
2 & \mbox{if} \;\;\frac{\mu'/2}{{\Delta'}^2 -1}=1  \\
4 & \mbox{if} \;\;\frac{\mu'/2}{{\Delta'}^2 -1}=-1
\end{cases}.
\end{equation}
Since $n_1$ and $n_2$ can not both be $4$ or $2$ simultaneously, we always have one order 2 saddle point and one order 4 saddle point in class II. The paths of  steepest descent  $C_{S^l_j}$ needed are  chosen in a way shown in Figure~\ref{fig.path_II_SC}. On the other hand, $\beta^{x,y,z}_1$ and $\beta^{x,y,z}_2$ are still given by Eq.\eqref{eq.beta1_SC} and Eq.\eqref{eq.beta2_SC}. Since the exponent of power-law decay of the order-4 saddle point is either  $5/4$ or $3/4$, which is always smaller then exponent of the order 2 saddle point ($5/2$ or $3/2$), we conclude that  
\begin{equation}\label{eq.result_II_SC}
\lambda_m(t) \sim \lambda_m(\infty) +
\begin{cases}
\frac{Q(t)}{t^{5/4}} & \mbox{if} \;\; \frac{\Delta'}{\Delta}=\frac{1\pm\mu'/2}{1\pm\mu/2}  \\
\frac{Q(t)}{t^{3/4}} & \mbox{if otherwise}
\end{cases},
\end{equation}
where $Q(t)$ contains one oscillating angular frequency: $\omega_1=|w(k_1)|=4|\mu'/2+1|$ when  $(\mu'/2)/({\Delta'}^2-1)=1$ or $\omega_2=|w(k_2)|=4|\mu'/2-1|$ when $(\mu'/2)/({\Delta'}^2-1)=-1$. Eq.\eqref{eq.result_II_SC} is confirmed by our numerical calculations shown in Figure~\ref{fig.result_II_SC}. In Figure~\ref{fig.result_II_SC}(a), the power-law behavior  with predicted  exponents (Eq.\eqref{eq.result_II_SC}) fits pretty well to the envelopes of the most important OPES $\lambda(t)$ for the two cases. On the other hand, the associated characteristic angular frequencies are also observed in the Fourier transformation of $\lambda(t)$ shown in Figure~\ref{fig.result_II_SC}(b).

\subsubsection{Class III: $|\frac{\mu'/2}{{\Delta'}^2 -1}|>1$}
For class III, in addition to the three order-2 saddle points $k_1$, $k_2$, one extra conjugate complex pair of  order-2 saddle point $k_3$ and  $k_3^*$ appears where $k_3$ and   $k_3^*$   are both solutions of $\cos^{-1}{[(\mu'/2)/({\Delta'}^2 -1)]}$. Those complex saddle point are irrelevant, which can be ignored. As a result  only the  two real  saddle points $k_1$ and $k_2$ have to be considered.

\begin{figure}
\center 
\includegraphics[width=12cm]{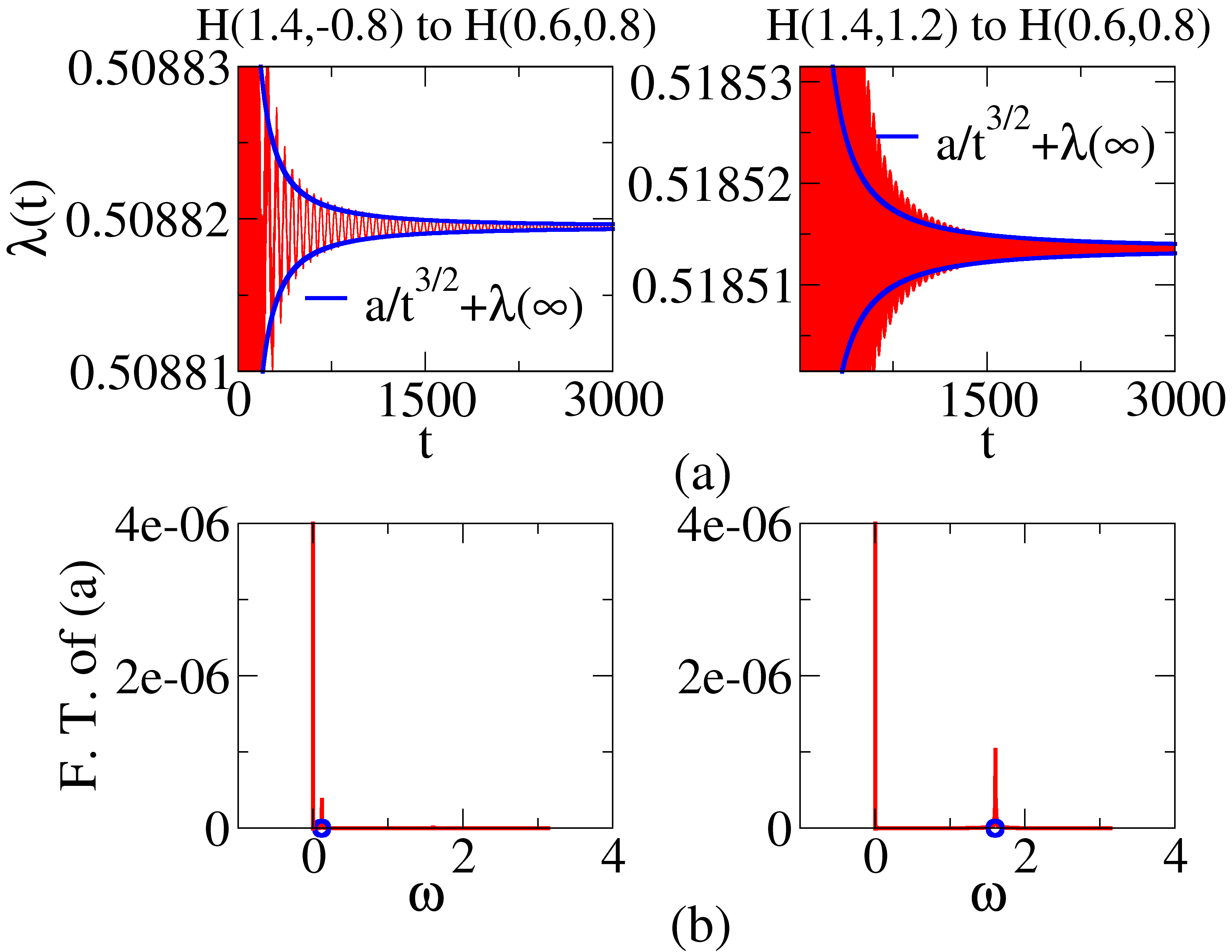}
\caption{(color online) 
The numerical calculations in class III when $\Delta'/\Delta=(1-\mu'/2)/(1-\mu/2)$ (left) and $\Delta'/\Delta=(1+\mu'/2)/(1+\mu/2)$ (right) with subsystem size $L=100$. (a) The most important OPES $\lambda(t)$ (red) and the power-law fitting $at^{-3/2}+\lambda(\infty)$ to the envelopes of $\lambda(t)$ (blue). (b) The Fourier transformation of (a). The characteristic angular frequency (open blue circle)  $\omega_1\approx 0.116815$ for left panel  and $\omega_2=1.6$ for the right panel are found. The peak at $\omega=0$ comes from the steady OPES $\lambda(\infty)$.
}
\label{fig.result_III_SC}
\end{figure}

$\beta^{x,y,z}_1$ and $\beta^{x,y,z}_2$ are the  same  as Eq.\eqref{eq.beta1_SC} and Eq.\eqref{eq.beta2_SC} with an extra condition: $\beta^z_1=\beta^z_2=\infty$ if $\Delta'=0$. However, this extra constraint does not influence the dominant exponent $\nu$ since the exponent for the extra condition is the maximal one.  
In this case $n_1=n_2=2$. Therefore we conclude that 
\begin{equation}\label{eq.result_III_SC}
\lambda_m(t) \sim \lambda_m(\infty) +\frac{Q(t)}{t^{3/2}},
\end{equation}
where $Q(t)$ contains oscillating frequencies $\omega_1$ when  $\Delta'/\Delta=(1-\mu'/2)/(1-\mu/2)$, $\omega_2$ when $\Delta'/\Delta=(1+\mu'/2)/(1+\mu/2)$, or $\omega_1$ and $\omega_2$ when $\Delta'/\Delta\neq(1\pm\mu'/2)/(1\pm\mu/2)$. The numerical calculations for the cases $\Delta'/\Delta=(1\pm\mu'/2)/(1\pm\mu/2)$ are shown in Figure~\ref{fig.result_III_SC}. In Figure~\ref{fig.result_III_SC}(a), the power-law function $a/t^{3/2}+\lambda(\infty)$ fits  the envelopes of the most important OPES $\lambda(t)$ fits very well. Furthermore, the corresponding characteristic angular frequencies are  also correctly found by taking Fourier transformation of Figure~\ref{fig.result_III_SC}(a) as shown in Figure~\ref{fig.result_III_SC}(b).

\begin{figure}[!h]
  \centering 
  \includegraphics[width=10cm]{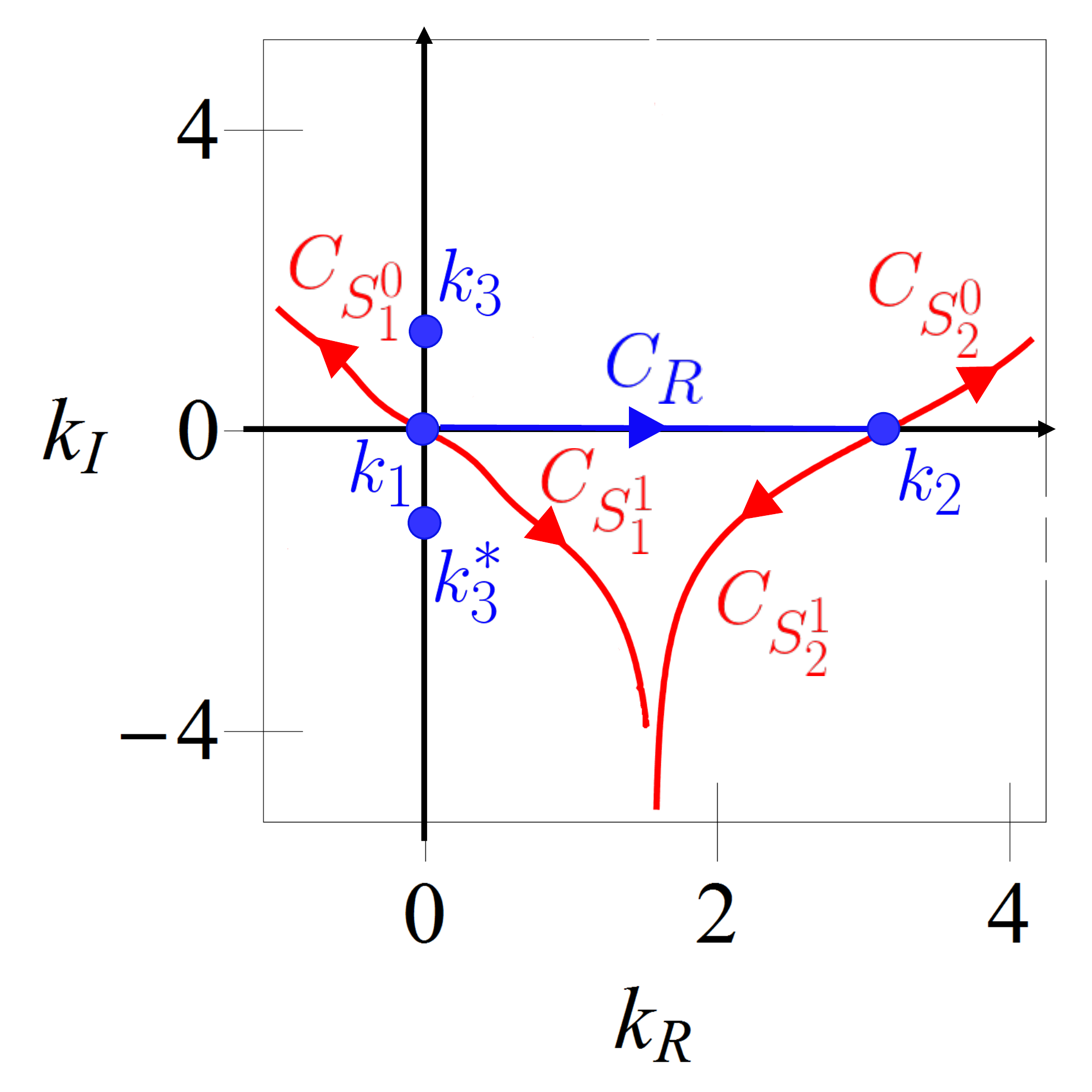}
  \caption
  {(color online)
  The typical steepest descent paths $C_{S^l_j}$ (red curves) started from the saddle points $k_j$ (blue dots) and the original integration path $C_R$ (blue line) in class III. We choose the path $C_S=C_{S^1_1}-C_{S^1_2}$ such that it does not passing through $k_3$ and $k_3^*$, meanwhile, still be a closed path in the complex plane. We use $(\mu'/2,\Delta')=(-0.7,0.8)$ for this figure.
  }
\label{fig.path_III_SC}
\end{figure}

\subsection{Discussion}
The reason why the steepest descent method works very  well can be understood as follows: 
In order to obtain  $k_j$, Eq.\eqref{eqA_def_kj} should be fulfilled. That means $\left.dR^\prime(k)/dk\right|_{k=k_j}=0$ since $\omega(k) \propto R^\prime(k)$. The saddle points $k_j$ can be interpreted as  the momentum of those quasiparticles with group velocity  $v_g\equiv dR^\prime(k)/dk$ equal to zero. According to Calabrese and Cardy\cite{Calabrese2005}, the system reaches its steady state when the quasiparticles hit the boundary between the subsystem and the environment, therefore the slowest quasiparticle dominates the asymptotic behavior. On the other hand, the steepest descent expansion will break down when the energy bands of $H'$ become extremely flat. The steepest decent method does not work well because the routes it choses are not much more important than the other routes due to the flat energy bands. The extreme case is that the energy bands are totally flat. The states only evolve with a phase shift $e^{-iEt}$. Therefore the steepest decent method totally fails.

The power-law decay exponents are found to be very different for the dimerized chains and 1-D $p$-wave superconductors. For dimerized chains $\nu=3/2$, which was  also found  for the transverse Ising model by Fagotti {\it et al.}\cite{Fagotti2013}. 
However, for 1-D $p$-wave superconductors, the decay exponents are not universal. 
This character essentially distinguishes dimerized chains and 1-D $p$-wave superconducting chains.

Especially for class I in $p$-wave superconducting chains, the decay exponent can be $1/2$, that is of one order magnitude slower that the decay exponent equal to $3/2$. 
As shown in the left panel of Figure~\ref{fig.result_I_SC}(a),  even to $t=30000$
 the system still not converges to its value of steady state. Therefore the decay exponents are very important for experimentalists to confirm their results, even they can not reach so long time limit in the experiments. 
On the other hand, in order to investigate the infinite time behavior numerically or experimentally, it is suggested by our finding that the parameters in class III should be chosen other than those in class I. If one is interested on choosing the parameters in class I due to some reality constraints, one should choose the parameters for the final Hamiltonian such that $(\mu'-\mu)=\Delta'(\mu'\Delta-\mu\Delta')$ to avoid the slowest power-law decay with exponent $1/2$. Just like what we find in dimerized chains, the power-law decays to the steady states of the OPES are not limited to the non-topologically steady states. As illustrated in Figure~\ref{fig.result_steady_topological}(b) for the parameters satisfied the upper one in Eq.\eqref{eq.result_II_SC}, the power-law decays with exponent $\nu=5/4$ to the topologically steady states are confirmed by our fittings.

\section{Conclusion and Outlook}\label{sec.SO}

\begin{figure}[!h]
  \centering 
  \includegraphics[width=12cm]{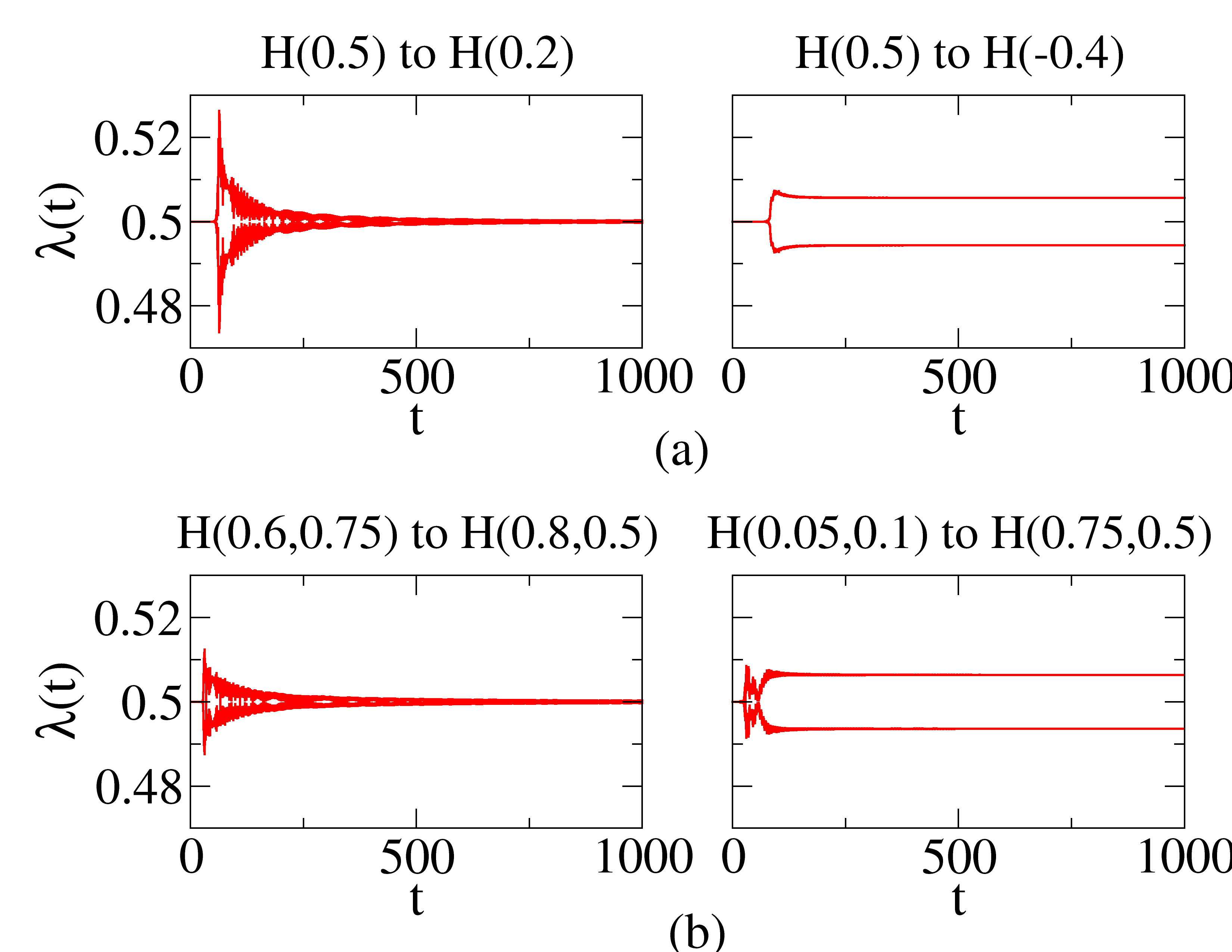}
  \caption
  {(color online)
  By showing the full evolutions of the most important OPES in  the right of Figure~\ref{fig.result_DC}(a), the left of Figure~\ref{fig.result_II_SC}(a) and Figure~\ref{fig.result_steady_topological}, one can observe the evolutions of the edge states of the OPES for (a) dimerized chains and (b) $p$-wave superconductors. 
  }
\label{fig.OPES}
\end{figure}

Starting from a one-dimensional topological system of an infinite length, either  a dimerized chain or a $p$-wave superconductor, and then measuring the entanglement spectra of a  subsystem $A$ with a length $L \gg 1$, one obtains  two edge states (or Majorana fermions for $p$-wave superconductors) with OPES $\lambda_m=1/2$. Those edge states  characterize the topological systems according to the bulk-edge correspondence of  entanglement measurements. After a sudden quench of the system at a certain time, we observe the evolution of those edge states (see Figure~\ref{fig.OPES}(a)) for dimerized chains  or Majorana fermions for $p$-wave superconductors(see Figure~\ref{fig.OPES}(b)). From the previous studies of the authors\cite{Chung2013,Chung2016}, some informations through the path of the evolution of OPES were  revealed. The sudden quench induces first the quasiparticle, however, those quasiparticles do not destroy the edge states until they meet each other in the middle of the system. That means, the two most important OPES keep their value as $1/2$ and then break up at the time when the quasiparticles reach the middle of the subsystem. Thus we obtain the quasiparticle velocity from the first plateau of $\lambda_m=1/2$ After that, they relax back to their values of steady states, either to the topological state ($\lambda_m=1/2$, left hand side)  or to the trivial state ($\lambda_m\neq 1/2$, right hand side) according the effective pseudo-magnetic field $\mathbf{S}_{\mathrm{eff}}$.

We already extracted some important  informations from the  early life and the final stage of the most important OPES. However, what happens in between was not clear at all before the present study has been made. This paper complements the details of  the life of edge states  and Majorana fermions between the plateau and the steady state. By using the steepest decent expansion, the long time behavior of the  most important OPES decays with a power-law exponent $\nu$ along with an oscillation characterized by some special  angular frequencies. The power-law decay exponents are totally different between dimerized chains and $p$-wave superconductors. While for dimerized chains, the decay exponents are always $3/2$, for $p$-wave superconductors the decay behavior is much richer. They can take different values such as $1/2,3/4,5/4,3/2,5/2$ depending on the parameter regime we chose. We also fit those analytical results with the numerics and find excellent agreements.  Actually, for power- law decay exponent $\nu=1/2$, the decay becomes very slow. It takes a very long time to reach the steady states. That is the reason why we have to do such kind of studies. Experimentally or numerically it is very difficult to obtain the infinite time behavior from the obtained data. Therefore one can use the knowledge about how the time-evolution of the system decays to its steady state to fit those data. From the intermediate stage of the time evolution, one extrapolates to the infinite time behavior of the systems. 

At this point we have to mention that the other OPES behave similarly like the most important OPES since the steepest decent method treats those OPES equally. We only mention the most important  OPES  because they  could be the signal of the topological systems and give us deep insight about the topological systems.

The power-law decay exponents are totally model dependent.
 They can be regarded as a ID card of the time evolutions for certain models.  
Till now there are only three models whose decay exponents $\nu$ are deciphered: the transversal Ising model by Fagotti {\it et. al.}\cite{Fagotti2013}, dimerized chains and $p$-wave superconductors by the authors.   Due to the fact that $\nu =3/2$ for transversal Ising models and dimerized chains, which is different from those exponents of $p$-wave superconductors, we can conclude that  transversal Ising models and dimerized chains are relatives. They actually can be transformed to each other by some unitary transformations. Naturally we ask a question: if these exponets can characterize the universality of different  models? To answer this question we need more studies on different models.


In the present paper, we concentrate on the decay properties  of OPES after a sudden  quench. As noted  in Ref.\cite{Fagotti2013}, the power-law convergence can be broken for other physical observables. To further check how robust the exponents and characteristic angular frequencies can be, it would be interesting to  analyse how other physical observables  converge. On the other hand, it is also interesting to ask whether the interaction will change our finding or not. However, the former problem needs more advance analytical techniques, which is beyond the content of this paper, while the latter one would need to overcome the numerical difficulty due to the linear growth of the entanglement as time increases. We would leave these  two possible issues for the future studies.

\ack
MCC and PCC acknowledge the NSC support under the contract Nos.
105-2112-M-005-009-MY3 and 104-2628-M-007-005-MY3, respectively.  

\appendix
\section{Steepest descent expansion}\label{sec.SDE}

In this work we often encounter real integral of the form
\begin{equation}\label{eqA_IR}
  \mathcal{I}_R(t)\equiv\int_{C_R}{f(k)e^{iw(k)t}dk},
\end{equation} 
where $C_R$ is the integration path along the real axis, $t$ is a real number, and $w(k)$ is a real function along the path $C_R$. Our interest is the asymptotic behavior of $\mathcal{I}_R(t)$ as $t\rightarrow\infty$. In the following we show the way to obtain the asymptotic behavior by applying the steepest descent expansion.

We will first sketch the whole procedure before going into the details. The first step is to form a close contour $C_O=C_R+C_S$ and apply the residue theorem to get
\begin{equation}\label{eqA_IR_Cauchy}
  \mathcal{I}_R(t) + \mathcal{I}_S(t) = \sum \mbox{residue of $f$ inside $C_O$},
\end{equation}
where
\begin{equation}\label{eqA_IS}
  \mathcal{I}_S(t)\equiv\int_{C_S}{f(k)e^{iw(k)t}dk}.
\end{equation} 
Since the right hand side of Eq.\eqref{eqA_IR_Cauchy} is time independent, how $\mathcal{I}_R(t)$ decays to its asymptotic value is solely determined by  $\mathcal{I}_S(t)$. To apply the method of steepest descent expansion, we first identify all the saddle points $k_j$, $j=1, 2, \cdots$, of $w(k)$, as well as their order $n_j$. For each saddle point $k_j$, we then find all the steepest descent paths $C_{S^l_j}$, $l=1, 2, \cdots$, that originate from $k_j$. Finally, we choose appropriate set of steepest descent paths $\{ C_{S^l_j} \}$ to form the path $C_S = \sum_{j, l} \pm C_{S^l_j}$ so that $\mathcal{I}_S(t)=\sum_{j,l} \pm \mathcal{I}_{S^l_j}(t)$, where
\begin{equation}\label{eqA_ISlj}
  \mathcal{I}_{S^l_j}(t)\equiv\int_{C_{S^l_j}}{f(k)e^{iw(k)t}dk}.
\end{equation} 
The sign in front of each steepest descent path is determined by the requirement that whether $C_{S^l_j}$ needs to be chosen in opposite direction or not. For each $\mathcal{I}_{S^l_j}$ we perform the integration and find the asymptotic behavior. We find that to the leading order it is always a power-law decay, accompanied with an oscillation. This implies that $\mathcal{I}_R(t)$ has the same asymptotic behavior, with exponent the same as the smallest exponent associated with $\mathcal{I}_{S^l_j}$. Similarly $\Delta G_{n,m}(t)$ will show the same asymptotic behavior, while its exponent is same as the smallest exponent associated with $\mathcal{I}^x(t)$, $\mathcal{I}^y(t)$, or $\mathcal{I}^z(t)$ defined in the main text.

\begin{figure}[!h]
\center
\includegraphics[width=12cm]{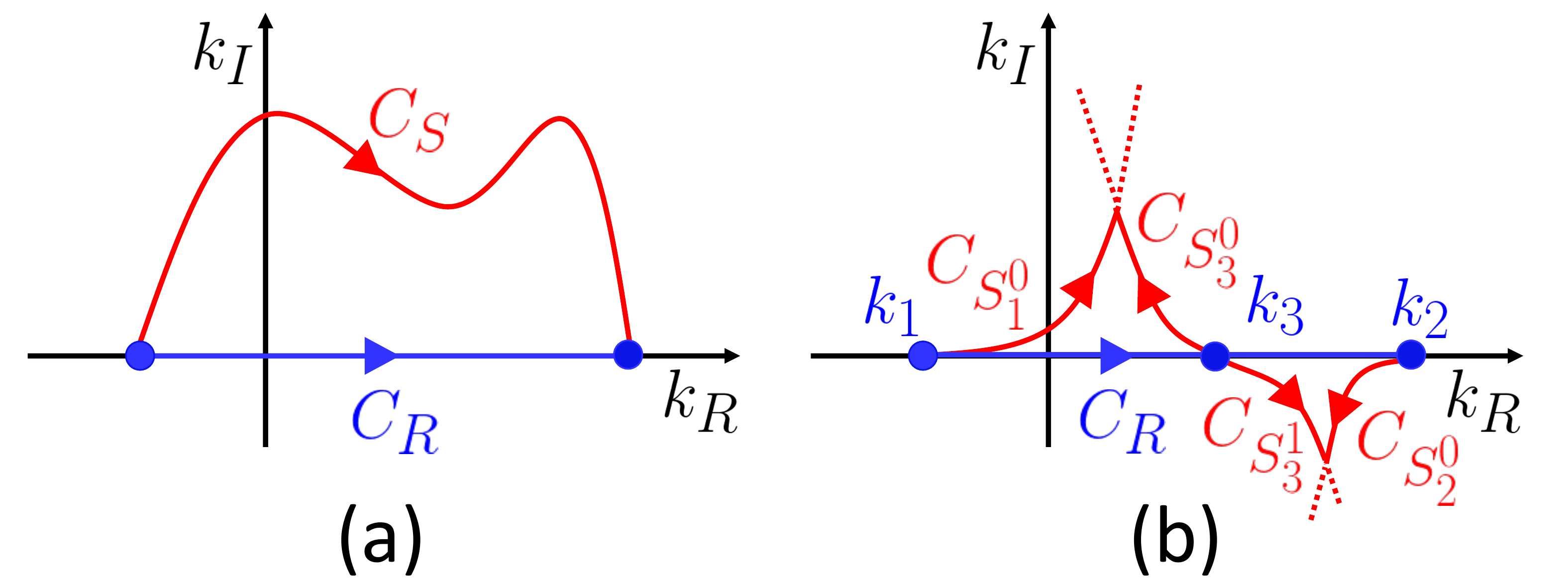}
\caption{(color online) 
Let the real and imaginary part of $k$ as $k_R$ and $k_I$. (a) A closed path $C_O=C_R-C_S$ can be constructed by the path $C_R$ along the real axis and path $C_S$ in the complex plane. (b) Given $k_1$ and $k_2$ are the boundary points and $k_3$ is a order $2$ saddle point, the path $C_S$ can be chosen as a combination of steepest descent paths $C_S=C_{S^0_1}-C_{S^0_3}+C_{S^1_3}-C_{S^0_2}$. Although the path $C_{S^l_j}$ should also contain the dashed part, we can only take the solid part as long as $t$ is large enough because $C_{S^l_j}$ is the steepest descent path.}
\label{figA_paths}
\end{figure}

Now we go through each step in detail. A saddle point is a point $k_j$ that some of the lower order derivatives of $w(k)$ are zero at $k_j$. Specifically, it is defined by
\begin{equation}\label{eqA_def_kj}
  w^{(l)}(k_j)
  \begin{cases}
    \neq 0 & l=0 \\
    = 0 & 1\leq l\leq n_j-1 \\
    \neq 0 & l=n_j \\
  \end{cases},
\end{equation}
where
\begin{equation*}
  \left.w^{(l)}(k_j)\equiv\frac{d^lw}{dk^l}\right|_{k=k_j}
\end{equation*}
is the $l$-th derivative of $\omega(k) $ at $k_j$. \textit{The order of the saddle point is defined as $n_j$}, the order of the first non-zero derivative of  of $\omega(k) $ at $k_j$. A steepest decent path $C_{S^l_j}$ starting from a saddle point $k_j$ is a path with the property that $-i(w(k)-w(k_j))$ is always real and positive along the path. In general, for a saddle point of order $n$, there will be $n$ steepest descent paths. To find the asymptotic behavior of $\mathcal{I}_{S^l_j}(t)$ as $t\rightarrow \infty$, we perform the change of variable $q\equiv -i(w(k)-w(k_j))$ and $\mathcal{I}_{S^l_j}(t)$ becomes

\begin{equation}
  \mathcal{I}_{S^l_j}(t)
  \equiv\int_{C_{S^l_j}}{f(k)e^{iw(k)t}dk}
  =e^{iw(k_j)t}\int_0^{\infty} \frac{-f(k)}{iw^{(1)}(k)} e^{-tq} dq.
\end{equation}
The upper limit is set to be the infinity since we are interested in the $t \rightarrow \infty$ behavior. We then expand $f(k)$ and $w^{(1)}(k)$ around $k_j$ to the next leading order and find
\begin{equation}
  w^{(1)}(k)=\frac{w^{(n_j)}(k_j)}{(n_j-1)!}\delta k^{n_j-1}\left( 
1+\frac{w^{(n_j+1)}(k_j)}{w^{(n_j)}(k_j)n_j} \delta k +O(\delta k^{2}) \right)
\end{equation}
and
\begin{equation}\label{eqA_f}
  f(k)=\frac{f^{(\beta_j-1)}(k_j)}{(\beta_j-1)!}\delta k^{\beta_j-1}+\frac{f^{(\beta_j)}(k_j)}{\beta_j!}\delta k^{\beta_j}+O(\delta k^{\beta_j+1}).
\end{equation}
We can see that $\beta_j-1$ shown in Eq.\eqref{eqA_f} is defined as the leading order of the Taylor expansion around the saddle point. The need to go to next leading order is that the leading order contribution from two $\mathcal{I}_{S^l_j}$ might cancel each other, as it will be shown later. To finish the change of variable, we need to relate $\delta k$ and $q$. To do this, we expand $q$ around $k_j$ up to the first non-zero term and find
\begin{equation}\label{eqA_q}
  q\approx -\frac{|w^{(n_j)}(k_j)|}{n_j!}|\delta k|^{n_j}e^{i(\alpha_j+n_j\theta_j)},
\end{equation}
where $\alpha_j$ and $\theta_j$ are the phase of $iw^{(n_j)}(k_j)$ and $\delta k$ respectively. For a steepest descent path, it is required that $q\equiv -i(w(k)-w(k_j))$ is always real and positive. It is evident that there are $n_j$ possible solutions to this requirement:
\begin{equation} \label{eqA_thetajl}
  \theta^l_j=\frac{(2l+1)\pi-\alpha_j}{n_j}, 
\end{equation}
where $l=0,1,\cdots,n_j-1$. As a result, there are $n_j$ steepest descent paths that are originated from an order $n_j$ saddle point. Now we can express $\delta k$ in terms of $q$
\begin{equation}\label{eqA_dk}
  \delta k\approx\left(\frac{qn_j!}{|w^{(n_j)}(k_j)|}\right)^{1/n_j} e^{i\theta^l_j},
\end{equation}
and perform the $q$ integration to obtain
\begin{equation}\label{eqA_ISlj_res}
  \mathcal{I}_{S^l_j}(t)=e^{iw(k_j)t} 
  \left[ 
    C^l_jt^{-\frac{\beta_j}{n_j}}
    +D^l_jt^{-\frac{\beta_j+1}{n_j}}
    +O\left(t^{-\frac{\beta_j+2}{n_j}}\right)
  \right].
\end{equation}
In the equation above
\begin{equation}
  C^l_j =
  \frac{f^{(\beta_j-1)}(k_j)}{n_j (\beta_j-1)!}
  \left( \frac{n_j!}{|w^{(n_j)}(k_j)|} \right)^{\frac{\beta_j}{n_j}}\Gamma\left(\frac{\beta_j}{n_j}\right) e^{i\beta_j\theta^l_j}
\end{equation}
and
\begin{equation}
  D^l_j =
  \left( \frac{f^{(\beta_j)}(k_j)}{n_j \beta_j!}-\frac{f^{(\beta_j-1)}(k_j)w^{(n_j+1)}(k_j)}{n_j^2 (\beta_j-1)!w^{(n_j)}(k_j)} \right)
  \left( \frac{n_j!}{|w^{(n_j)}(k_j)|} \right)^{\frac{\beta_j+1}{n_j}}\Gamma\left(\frac{\beta_j+1}{n_j}\right)e^{i(\beta_j+1)\theta^l_j},
\end{equation}
where $\Gamma(x)$ is the gamma function.

All the integrals in this paper, however, have two additional properties: 
(i) the two boundary points are always saddle points, and 
(ii) we can always choose saddle points on the real axis and use their corresponding steepest descent paths to form the contour. From Eq.\eqref{eqA_ISlj_res} we find that to the leading order, asymptotically $\mathcal{I}_{S^l_j}(t)$ behaves as
 
\begin{equation}\label{eqA_ISlj_leading}
  \lim_{t\rightarrow\infty}\mathcal{I}_{S^l_j}(t) \sim e^{iw(k_j)t}\frac{C^l_j}{t^{\beta_j/n_j}},
\end{equation}
which is a power-law decay with exponent $\beta_j/n_j$, accompanied  with an oscillation. The characteristic angular frequency of the oscillation reads 
\begin{equation}\label{eqA_omega}
  \omega_j=|w(k_j)|.
\end{equation}
Note that the angular frequency is independent of the paths as we can see from Eq.\eqref{eqA_ISlj_res}.
Moreover, the  Riemann-Lebesgue lemma guarantees $I_R(t)\rightarrow 0$ as $t\rightarrow\infty$ as long as $f(k)$ and $w(k)$ are  smooth and well-behaved functions along the path $C_R$. Hence, the time independent residue term in Eq.\eqref{eqA_IR_Cauchy} has to vanish at any time $t$ and we obtain the integral Eq.\eqref{eqA_IR} as
\begin{equation}\label{eqA_IR_ours}
  \mathcal{I}_R(t)=\sum_{l,j}{\pm \mathcal{I}_{S^l_j}(t)},
\end{equation}
From Eq.\eqref{eqA_IR_ours} and the Eq.\eqref{eqA_ISlj_leading}, we finally conclude that the asymptotic behavior of $\mathcal{I}_{R}(t)$ as $t\rightarrow\infty$ is given by 
\begin{equation}\label{eqA_IR_asym}
  \mathcal{I}_R(t)\sim \frac{1}{t^{\nu}} \sideset{}{'} \sum_{l,j}a^l_je^{i\omega_jt},
\end{equation}
where $\nu$ is the minimum of $\beta_j/n_j$, $a^l_j$ is a constant and the  summation $\sum^\prime_{l,j}$ only sum over those $l,j$ with  the constraint: $\nu=\beta_j/n_j$. 

When $\beta_s = n_s$, however,  Eq.\eqref{eqA_IR_asym} needs to be modified. This happens, for example, for the case of class I of $p$-wave superconducting chains. With this condition, one has $e^{i\beta_s\theta^l_s} = e^{(2\l+1)\pi-\alpha_s} =-e^{-i\alpha_s}$ as seen from Eq.\eqref{eqA_thetajl}. When two steepest descent paths originated from the same saddle point with $\beta_s = n_s$ are both chosen, the two coefficients of the leading term are the same and the leading order contributions cancel each other due to the fact that one has to chose a path in opposite direction and pick up a minus sign. Therefore the asymptotic of $-\mathcal{I}_{S^p_s}(t)+\mathcal{I}_{S^{p'}_s}(t)$ is dominated by the sub-leading terms. With some algebra one finds

\begin{equation}\label{eqA_IR_asym_spe}
  -\mathcal{I}_{S^p_s}(t)+\mathcal{I}_{S^{p'}_s}(t)\sim \left(e^{i\frac{(2p'+1)\pi}{n_s}}-e^{i\frac{(2p+1)\pi}{n_s}}\right)\frac{E_s}{t^{\nu_s}}e^{i\omega_st},
\end{equation}
where $\nu_s=1+1/n_s$ and
\begin{equation*}
  E_s=-\frac{B_s}{n_s}\left( \frac{n_s!}{|w^{(n_s)}(k_s)|} \right)^{1+\frac{1}{n_s}}\Gamma\left(1+\frac{1}{n_s}\right)e^{-i\alpha_s(1+1/n_s)}.
\end{equation*}

\section*{References}
\bibliographystyle{iopart-num}
\bibliography{ref}
\end{document}